\documentclass[prb,aps,amssym,nofootinbib,floatfix,eqsecnum,notitlepage,showpacs]{revtex4-1} 
\usepackage{epsfig,natbib}
\usepackage{graphicx}
\newcommand{\beq}{\begin{equation}}
\newcommand{\eneq}{\end{equation}}
\newcommand{\be}{\begin{equation}}
\newcommand{\ee}{\end{equation}}
\newcommand{\bea}{\begin{eqnarray}}
\newcommand{\eea}{\end{eqnarray}}

\begin{document}
\title{The dc Josephson current in a long multi-channel quantum wire}
\author{Domenico Giuliano$^{1,3}$ and Ian Affleck$^{2}$}
 \affiliation{$^{1}$
Dipartimento di Fisica, Universit\`a della Calabria Arcavacata di Rende I-87036, Cosenza, Italy
and
I.N.F.N., Gruppo collegato di Cosenza, Arcavacata di Rende I-87036, Cosenza, Italy\\
$^{2}$ Department of Physics and Astronomy, University of British 
Columbia, Vancouver, B.C., Canada, V6T 1Z1\\
$^{3}$ CNR-SPIN, Monte S.Angelo – via Cinthia, I-80126 Napoli, Italy}
\date{\today}

\begin{abstract}
The dc Josephson current across a multi-channel SNS junction is computed
by summing  contributions from sub-gap Andreev bound states, as well as
from continuum states propagating within the superconducting leads. We show that, 
in a   long multi-channel SNS-junction, at low temperatures, all  these contributions
add up, so that the current  can be entirely expressed in terms
of single-particle  normal- and Andreev reflection amplitudes at the Fermi level at 
both SN interfaces. Our derivation  applies to a 
generic number of channels in the normal region and/or in the superconducting leads,
without assumptions about scattering processes at the SN interfaces:
if the  channels within the 
central region have the same dispersion relation, it leads to simple 
analytical formulas for the current at low temperatures;
if the channels within the central region have different dispersion relations, 
it allows for expressing the current in terms of a simple integral   involving 
only scattering amplitudes at the Fermi level.
Our result motivates using a low energy effective boundary
Hamiltonian formalism for computing the current, which is crucial for treating 
Luttinger liquid interaction effects. 
\end{abstract}


\maketitle

\section{Introduction}
\label{intro}

The dc Josephson current  \cite{Andreec} flowing at zero voltage bias 
across an SNS-junction at temperature $T$ as 
a consequence of an applied phase difference $\chi$ between the superconducting leads, 
is generally obtained \cite{anderson} by taking the
derivative of the system free   energy $F$ with respect to   $\chi$,  that is
$I [ \chi ; T]  = 2 e \frac{ d  F }{ d \chi }  $.
Using  Bardeed-Cooper-Schrieffer (BCS) approximation
for the leads and ignoring interactions within the normal region, assuming 
spin rotational symmetry for the whole system, the free energy $F$
 is simply obtained by summing over all the individual
single-quasiparticle energies $E_n$, so that the current is given by  \cite{anderson}

\beq
I [ \chi ; T ]  = 2e\sum_ nf(E_n) {dE_n\over d\chi}
\:\:\:\: , 
\label{current.2}
\eneq
\noindent
with  $f ( E )=1/[e^{E/T}+1]$ 
being the Fermi distribution function which,   at $T=0$,  gives the zero-temperature 
current $I  [ \chi ; T=0 ] = 2e\sum_ { E_n \leq 0}  {dE_n\over d\chi}$.
The factor of 2 in Eq. (\ref{current.2}) accounts for the spin degeneracy of
each level, due to the spin rotational symmetry of the system.\cite{been.1,aff_giu_new}
In general, to compute $ I [ \chi ; T ]$  one has to pertinently sum over contributions
from both sub-gap Andreev bound states (ABS's),\cite{andreev} localized in the central normal region, 
with  wavefunctions exponentially decaying within the superconducting leads,  
as well as from propagating scattering states (SS's), with energy $ | E | > \Delta$, $\Delta$ being 
the superconducting gap.\cite{kulik,ishi} Accurately summing over all types of states 
is, in general, quite hard, due to the delicate cancellation between various contributions, 
yielding a small final result from differences of very large terms.\cite{bardeen}

In Ref. [\onlinecite{giu_af}], based on an adapted  version of the formalism developed in
Refs.[\onlinecite{been.0,furus}], we rewrite Eq. (\ref{current.2}) as a  contour integral in 
the complex energy plane, explicitly involving the determinant of the 
analytically continued  $S$-matrix, 
from which we show  that, at low temperatures,  the formula for 
the dc Josephson current is greatly simplified in the long junction limit. In particular, 
we prove that, for a long junction and at low temperatures, $I  [ \chi ; T ]$ 
depends only on data at the Fermi level, 
namely, on the single-particle normal- and Andreev-reflection amplitudes at
the SN interfaces. Specifically, in [\onlinecite{giu_af}] we consider a one-dimensional model for the SNS-junction, 
with just one active channel, within both the central region, and the superconducting leads,
both in the continuum formulation (``Blonder-Tinkham-Klapwijk (BTK) model'' \cite{btk}), as 
well as in a tight-binding version of the model, such as the one discussed in 
Ref. [\onlinecite{ACZ}].
Nevertheless, ballistic SNS junctions realized with point contacts  between superconducting leads 
(``superconducting quantum contacts''), \cite{sqpc,been.0} as well as by connecting, for instance, a carbon nanotube
to two superconductors,\cite{kouwev} are typically characterized by several open one-dimensional
channels, both within the leads and in the central region.  To keep in touch with 
such realistic models of ballistic SNS junctions, in this paper we 
discuss the generalization of the main results of 
[\onlinecite{giu_af}] to a long SNS junction at low temperatures, with an arbitrary number of
open channels $N_L$ and $N_R$ within the left-hand  (L)  and 
the right-hand superconducting lead (R), respectively, and a generic number $K$ of (noninteracting) 
open  electronic channels within  the central region C. In particular, on providing 
explicit applications of our approach to specific model-calculations of $ I [ \chi ; T ]$,
we show that,  the simple closed-form formula for  $ I [ \chi ; T ]$ 
($I  [ \chi ; T=0 ]$) given in Eq. (3) of [\onlinecite{giu_af}] 
takes a nice generalization to the multi-channel SNS junction
in the case of equivalent channels within C, that is, in the case 
in which all the channels within C are characterized by the same
dispersion relation (but not necessarely by the same 
tunneling amplitudes with  the leads). Even when this last condition is not met, it
is possible to write compact formulas for $ I [ \chi ; T ]$ at low temperatures, 
which eventually allow for a straightforward   calculation of the current.
 
  In treating   ballistic multi-channel SNS junctions, relevant results
have been obtained by using a quasiclassical approach based on the Eilenberger
equations written for the slowly varying (on atomic distances) part of the 
Matsubara-Green functions,\cite{zaikin_1} in which the scattering at SN interfaces
is accounted for by means of simple linear conditions, rather than 
the ``standard'' Zaitsev boundary conditions.
\cite{zaitsev} Such an approach has revealed itself to be quite effective in
computing the dc Josephson current in a variety of physically relevant
situations, such as Josephson junctions with a series of insulating barriers,\cite{gala_zaikin} 
SHS junctions, where H is a ``half-metal'', that is, fully 
spin polarized materials acting
as an insulator for electrons with one of the two spin directions,\cite{gala_zaikin_2} junctions 
realized with spin-active SN-interfaces,\cite{gala_zaikin_3} or junctions 
realized with single- or multi-layer graphene contacted with two superconducting electrodes.\cite{takane}

As we   outline in Appendix \ref{phi_omchi}, in the long-junction limit, 
the results of  Ref.[\onlinecite{gala_zaikin}] can be recovered from our
simple formulas, in 
the limit in which one  neglects scattering between different channels at
the SN-interfaces. Thus, while being consistent with 
the well-grounded method based on Eilenberger
equations, our approach consitutes a remarkable simplification
of the technique  of Ref.[\onlinecite{gala_zaikin}], as  it 
provides  an explicit formula for  the current 
at low temperatures in a generic long ballistic multi-channel SNS
junction, without going though a limiting procedure of complex formulas. Moreover, the fact that we  explicitly prove the
cancellation between contributions to $I [ \chi ; T ]$ from 
finite-energy states,  
 motivates resorting to a simplified model calculation, in which 
the superconducting leads are integrated out and traded for a pertinent boundary interaction
Hamiltonian, only involving the single-electron field operators at the endpoints of the central region. 
This result is important, as it provides an effective method for including  interaction effects in the 
normal region, based on boundary conformal field theory techniques,\cite{mastone,ACZ,Titov} and can be
readily applied to study, for instance, the signature on the dc Josephson current 
of the emergence of nontrivial fixed points in junctions involving topological 
superconductors,\cite{giu_af2}  quantum Josephson junction networks,\cite{qjj} etc. 

It is worth remarking that, while, in order to present our technique, 
throughout this paper we work with a model Hamiltonian in which the leads 
are pictured as one-dimensional s-wave superconductors and the whole SNS-Hamiltonian
is $SU(2)$-invariant, so that spin is conserved in a single-quasiparticle 
scattering at the SN-interfaces,  with pertinent modifications to
the model Hamiltonian used for the calculations, our derivation is  expected to be
effective in providing a reliable long-junction limit of the dc Josephson current 
in the system studied, for instance,  in [\onlinecite{gala_zaikin_2},\onlinecite{gala_zaikin_3}].  
 
The paper is organized as follows:

\begin{itemize}

\item{ In Section \ref{multi_cha}, we implement a pertinent version of the 
$S$-matrix approach, to  derive the general formula for the dc Josephson current 
across a multi-channel SNS junction.}

\item{In Section \ref{long_junction} we compute the dc Josephson current at low 
temperatures 
across a long multi-channel SNS junction. We show that, to leading order in  $\ell^{-1}$,
the current is fully determined only by scattering amplitudes at the Fermi level. In particular, 
in the case of $K$ equivalent channels within C, the current can be presented in a simple 
closed-form formula, in terms of the roots of an algebraic equation of the form 
${\cal P} ( u ; \chi ) = 0$, with ${\cal P} ( u ; \chi )$ being, at fixed $\chi$, 
a $2K$-degree polynomial of the (complex)  
unknown $u$. In the case  of $K$ inequivalent channels within C, 
at low temperatures the leading contribution to the current in  $\ell^{-1}$ 
can be recast in an integral  formula that can be easily computed numerically.}

\item{Section \ref{concl} contains our conclusions.}

\item{In the appendices, we provide mathematical details of our derivation.}
\end{itemize}

\section{The dc Josephson current for a multi-channel   SNS junction}
\label{multi_cha}

In this section, based on a minimal set of reasonable assumptions, we provide a general formula 
for $ I  [ \chi ; T=0 ]$ and eventually discuss the extension of 
the result to $ I [ \chi ; T ]$. 

To simplify the derivation of  the general formula for $I  [ \chi ; T=0 ]$, we 
assume that, while the dispersion relation within the superconducting lead
can be different for different channels, the number of channels in the two leads
is the same, that is,  $N_L = N_R \equiv N$. In addition, we assume that the superconducting 
order parameter is the same for each channel (see Appendix \ref{models} for a detailed
review of the simplifying assumptions.)  As discussed above, 
while, as a model calculation, we consider 
the case of $s$-wave superconducting leads, pictured as one-dimensional
superconductors described by the model Hamiltonian introduced in 
[\onlinecite{btk}], but   our derivation is expected to apply  equally well, for instance, to 
lattice models,\cite{ACZ}, to SHS-junctions \cite{gala_zaikin_2},
to junctions with spin-active interfaces \cite{gala_zaikin_3}, 
or to the case in which the leads are realized with   
superconductors with unconventional pairing, which  has recently become of  
great relevance to engineering SN-interfaces hosting localized 
Majorana fermions.\cite{kita_1,lutch_1,oreg,lut_majo,procolo_1,procolo_2}

\begin{figure}
\includegraphics*[width=.8\linewidth]{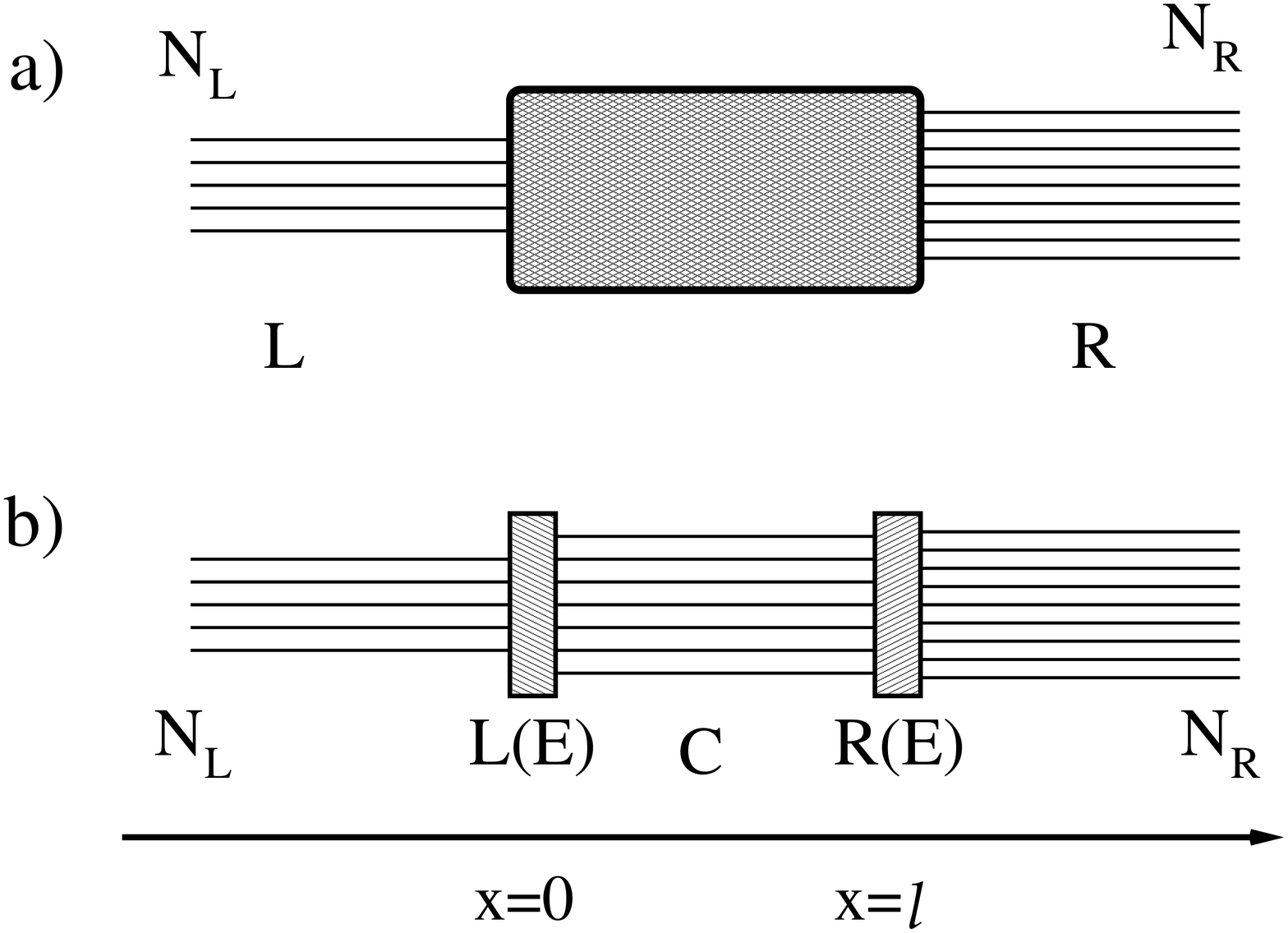}
\caption{{\bf a):} Sketch of a generic SNS junction with $N_L$ open channels within 
the left-hand lead and $N_R$ open channels within the right-hand lead. The $S$-matrix
is fully determined by the scattering processes that happen in the region
corresponding to the central box of the figure. 
{\bf b):} Sketch of a long multi-channel SNS junction with $N_L$ open channels within 
the left-hand lead and $N_R$ open channels within the right-hand lead. The dashed boxes
represent the transmission matrices $L(E)$, connecting the left-hand lead to the central region 
C, and $R(E)$, connecting C to the right-hand lead.} \label{fig_multi0}
\end{figure}

In Fig. \ref{fig_multi0} {\bf a)}, we provide a sketch of a generic multi-channel  
SNS junction. The corresponding Hamiltonian is given in Eqs. (\ref{newvel.01}), (\ref{newvel.02}) and (\ref{sss.b1}). 
Note that it has s-wave pairing and SU(2) spin symmetry. 
Any superconducting region $r$ of the junction is described by a model Hamiltonian of the form

\begin{eqnarray}
 && H_r - \mu {\cal N}_r  = \int_{ x \in r}  \: d x \:  \: \sum_{ \lambda= 1}^{N_r} 
 \biggl\{ \sum_\sigma   
 \Psi_{ r , \lambda , \sigma}^\dagger 
 ( x ) h_{0,\lambda} ( x )  \Psi_{r , \lambda , \sigma}  ( x ) \nonumber \\
 &+& \Delta  e^{  \frac{i}{2} \chi_r }  \Psi_{r , \lambda , \uparrow}  ( x ) 
 \Psi_{r , \lambda , \downarrow}  ( x ) + 
 \Delta  e^{ -  \frac{i}{2} \chi_r }  \Psi_{r , \lambda ,\downarrow }^\dagger   ( x ) 
 \Psi_{r , \lambda , \uparrow }^\dagger   ( x ) \biggr\} 
 \;\;\;\; , \nonumber \\
 \label{newvel.0x}
\end{eqnarray}
\noindent
with $N_r$ being the number of active channels and ${\cal N}_r$ being the total 
particle number within $r$, $h_{0,\lambda} ( x )$
being the normal Hamiltonian of $r$, $\Delta$ being the superconducting order paramenter
and $\chi_r$ being the corresponding phase. $\Psi_{r , \lambda  , \sigma}  ( x )$ is
the single-electron field operator for a particle in channel $\lambda$ with spin $\sigma$. 
The key quantity required to compute the dc Josephson current 
across a junction as such is the $S$-matrix for single quasiparticle states. Indeed, 
in general it can be shown that  all the contribution to the dc Josephson current add up to
an integral formula which only depends on the determinant of $S$.\cite{ishi,been.0,furus,giu_af}
In Appendix \ref{models_details} we discuss in detail the derivation of the single-particle wavefunctions 
in the leads from the  BDG  equations  and 
the corresponding definition of the $S$-matrix. 
Since $S$ is defined in terms of the ``asymptotic'' (that is, far enough from 
the central region) behavior of the wave functions, the formulas we derive  in this section 
do not rely on any specific assumptions concerning C and the SN-interfaces, such as the 
ones we will introduce to discuss the long-junction limit, and   hold independently of 
the specific behavior of the superconducting gap at the SN-interfaces and/or of the 
particular form of the Hamiltonian within C.

\noindent

In Ref. [\onlinecite{giu_af}], the 
key step to systematically work out the formula for $I [ \chi ; T ]$ in the long junction limit 
in the single-channel case was the 
possibility of expressing the determinant of the $S$-matrix at fixed $E$ and $\chi$,  
${\rm det} [ S ( E ; \chi ) ]$ as

\beq
{\rm det} [ S  (E ; \chi ) ]  = \frac{ {\cal F} [ E ; \chi ] }{ {\cal G} [ E ; \chi ] }  
\:\:\:\: ,
\label{mc.20}
\eneq
\noindent
with  ${\cal F} [ E ; \chi ]$  and ${\cal G} [ E ; \chi ]$
in Eq. (\ref{mc.20}) being functions of $E$ in the complex $E$-plane
which  we choose to obey the following properties (generally met in physically
relevant models):\cite{giu_af} \\
{\it i)} They are always finite   for   finite $E$. This can be easily achieved by 
shifting poles of ${\cal G}$ into zeroes of ${\cal F}$ and vice versa; \\
{\it ii)} They have no common zeroes. [Possible common zeroes (e.g. $E_0$), could 
always be cancelled by a redefinition: ${\cal F} ( E ; \chi )\to 
{\cal F} [ E ; \chi ]/(E-E_0)$, ${\cal G} [ E ; \chi ]\to {\cal G} [ E ; \chi ]/(E-E_0)$, 
without changing Eq. (\ref{mc.20})].\\
{\it iii)} ${\cal F} [ E ; \chi ]={\cal G}^* [ E ; \chi ]$. 
Here this equation refers to complex conjugating {\it the function} without complex 
conjugating its argument, $E$. This condition is consistent with 
the requirement that $ | {\rm det} [ S ] | = 1$ for scattering states.\\
{\it iv)}  ${\cal G} [ E ; \chi ]$ can be defined to have  branch cuts along the real $E$-axis,
corresponding to the nonzero density of scattering states in the leads. This is due to the fact that
${\cal G} [ E ; \chi ]$ depends on $E$ via the particle and hole momenta $\beta_p $ and $\beta_h$ and that they
become double-valued functions of $E$, for $| E | > \Delta$.\\
{\it v)} $\partial_\chi \ln {\cal G}[ E ; \chi ]   $ vanishes rapidily at $|E|\to \infty$
 along any ray not parallel to the real axis.\\
{\it vi)} ${\cal G} ( E ; \chi )$  is real in the bound state region: the real axis with $- \Delta 
\leq E \leq \Delta$. 

\begin{figure}
\includegraphics*[width=1.05\linewidth]{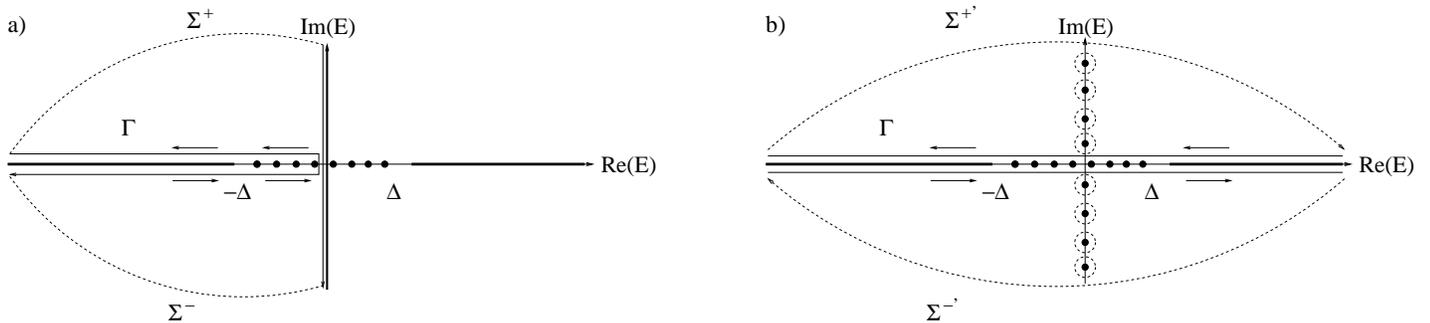}
\caption{{\bf a):} Deformation of the integration path in the energy plane to the imaginary axis
to compute $I  [ \chi ; T=0 ]$. The solid lines on the real axis correspond to 
branch cuts in $ G [ E ; \chi  ]$, the solid dots corresponds to poles, that is, 
to Andreev bound states. The integrals over the arcs $\Sigma^\pm$ go to zero, as 
the radius of the arcs goes to $\infty$, as a consequence of the hypotheses on
the behavior of the function $G [ E ; \chi]$; 
{\bf b):} Deformation of the integration path in the energy plane to the imaginary axis
to compute $I  [ \chi ; T]$ at finite $T$: the integral over the imaginary axis is 
substituted by a sum over the Fermionic Matsubara frequencies, as it can be shown by
deforming the integration path to a series of small circles encircling the 
points on the imaginary axis at $i \omega_\nu$. } \label{fig_multi1}
\end{figure}

Once the above conditions are met, from Eq. (\ref{mc.20}), by  deforming the integration path 
as displayed in FIg. \ref{fig_multi1} {\bf a)},  one can first of all 
show \cite{giu_af} that $I  [ \chi ; T=0 ]$ can be written in terms
of just one integral over the imaginary axis as
 
\beq
I  [ \chi ; T=0 ]= \frac{2e}{2 \pi} \; \int_{- \infty}^\infty \; d \omega \: 
\partial_\chi \ln {\cal G} [ i \omega ; \chi ] 
\:\:\:\: . 
\label{mc.24}
\eneq
\noindent
At variance, as sketched  in Fig. \ref{fig_multi1} {\bf b)}, at finite-$T$ the deformation of 
the integration path in the energy plane yields a sum over the 
fermionic Matsubara frequencies $\omega_\nu = 2 \pi T \left( \nu + \frac{1}{2} \right)$, with 
$\nu$ being a relative integer, so that one obtains \cite{giu_af}

\beq
I [ \chi ; T ] = 2e T \; \sum_\nu  \:  
\partial_\chi \ln {\cal G} [ i \omega_\nu ; \chi ] 
\:\:\:\: .
\label{mc.24ft}
\eneq
\noindent
For a single-channel junction, the function  ${\cal G} [ E ; \chi ]$ is
given by \cite{giu_af} 
 
\beq
  {\cal G} [ E ; \chi ]  =  [ M ( E ; \chi ) ]_{2,2}  [ M ( E ; \chi ) ]_{4,4} 
  - [ M ( E ; \chi ) ]_{2,4} [ M ( E ; \chi ) ]_{4,2}
 \;\;\;\; , 
 \label{nov.1}
 \eneq
 \noindent
 where $[ M ( E ; \chi ) ]_{i , j }$ are the matrix elements of the transmission matrix 
 $M ( E ; \chi )$, defined in the general case in Eq. (\ref{mc.6}) of Appendix \ref{models_details}.
 Eqs.(\ref{mc.24ft},\ref{mc.24}) are formally equivalent to 
Eq. (\ref{current.2}) and to its zero-temperature limit respectively, so, they 
are exact formulas, independently of the details of the SNS junction.  To generalize 
them to the multi-channel case, we first of all introduce a pertinent 
labeling of the transmission matrix element, namely, we label 
each matrix elements $[ M ( E ; \chi )]_{ ( j , \lambda) , (j' , \lambda' ) }$ with 
the pair of indices $(j, \lambda )$ and $(j' , \lambda' )$.   
$ \lambda  , \lambda' ( =  1 , \ldots , N)$  are the channel indices, 
while $j , j' = 1 , \ldots , 4$ label the forward/backward-propagating particle/hole-solutions, 
exactly as in the single-channel case. In order to generalize  Eq. (\ref{mc.20})
to the multi-channel case, we have to derive the generalized  
${\cal F} [ E ; \chi ], {\cal G} [ E ; \chi ]$ functions. This is done in 
Appendix \ref{proof}, where we prove that one gets

\begin{eqnarray}
  {\cal F} [ E ; \chi ] &=& {\rm det}  [ M^A  ( E ; \chi ) ] \nonumber \\
   {\cal G} [ E ; \chi ] &=& {\rm det}  [ M^B   ( E ; \chi )] 
 \:\:\:\: , 
 \label{mc.25}
\end{eqnarray}
\noindent
with $M^A , M^B$ in Eq. (\ref{mc.25}) being $4 N \times 4 N$-matrices  that 
are given by

\beq
[ M^A ( E ; \chi ) ]_{(j , \lambda) , ( j' , \lambda' ) } = ( \delta_{j, 1}+\delta_{j, 3} ) [ M ( E ; \chi ) ]_{(j , \lambda) , ( j' , \lambda' ) }  
+  ( \delta_{j, 2}+\delta_{j, 4} )\delta_{jj'}
 \delta_{\lambda , \lambda'}
\:\:\:\: , 
\label{mc.18}
\eneq
\noindent
and

\beq
[ M^B  ( E ; \chi ) ]_{(j , \lambda) , ( j' , \lambda' ) }   ( \delta_{j , 2} + 
\delta_{j , 4}  )  [ M ( E ; \chi ) ]_{(j , \lambda) , ( j' , \lambda' ) }+  
( \delta_{j, 1}+\delta_{j, 3})\delta_{jj'}
 \delta_{\lambda , \lambda'}
\:\:\:\: . 
\label{mc.19}
\eneq
\noindent
Eqs.(\ref{mc.20},\ref{mc.25},\ref{mc.18},\ref{mc.19}) encode the key
result  of this section. Based on these equations, in the following, we discuss in detail the simplifications that  occur 
to  Eqs.(\ref{mc.24},\ref{mc.24ft})  
in the long junction limit,  also providing a few explicit model calculations of 
the dc Josephson current. For the sake of the presentation, we will separately 
discuss the ``symmetric'' case, in which the $K$ channels within C show the 
same dispersion relation, and the case in which such a symmetry is lacking. 
In fact, as we will show, while in both cases the current only depends on 
scattering amplitudes at the Fermi levels,  in the former case it is also possible 
to provide a simple closed-form formula for the 
current, which is exact to leading order in the inverse length of the junction.

\noindent

\section{Josephson current across a long multi-channel SNS junction}
\label{long_junction}

 In order to discuss the  long-junction limit, we assume that the system exhibits sharp interfaces between
the leads and the central region,\cite{note_rigid} as we sketch in FIg. \ref{fig_multi0} {\bf b)}.
In particular, we assume that  the central region C runs from $x = 0$ to $x = \ell$ (as from now on 
$\ell$ will be the key variable of the expansion we perform,  we will explicitly display it among 
the argument of the various functions.) Thus, as 
we are considering a ballistic SNS junction,  the long-junction limit is defined by  $E_{\rm Th} \ll \Delta$ 
with the  Thouless energy  $E_{\rm Th} \sim v / \ell$, $v$ being of the 
order  of the Fermi velocity within C. As we discuss in detail in Appendix \ref{models_long}, 
the transmission  matrix $M ( E ; \chi ; \ell ) $ can then be written in a factorized form as 

\beq
M ( E ; \chi ; \ell ) = R ( E  ; \chi ) \cdot M^C ( E ; \ell) \cdot L ( E ; \chi  ) 
\;\;\;\; , 
\label{mm.1}
\eneq
\noindent
with $L(E ; \chi )$ ($R(E ; \chi )$)  being the $4N \times 4K$ ($4K \times 4N$) transmission matrix 
at the left- (right-) hand interface, and $M^C ( E ; \ell )$ being the transmission matrix of the central region.
For a ballistic junction, we then obtain  
$[M^C ( E ; \chi )]_{(j,\rho) ; (j' , \rho' )}^C (E) = e^{ i \alpha_{j , \rho} \ell} \delta_{ j , j'}
\delta_{ \lambda , \lambda'}$ and $\alpha_{1(2) , \rho} = \pm \alpha_{ p , \rho}$,
$\alpha_{3(4) , \rho} = \mp \alpha_{ h , \rho}$ ($\rho , \rho' = 1 , \ldots K$), with 
the energy-$E$ particle- and hole-momenta within channel $\rho$, $\alpha_{ p , \rho} ,\alpha_{ h , \rho}$,
defined in Eq. (\ref{centr.3}) of Appendix \ref{models_long}. From Eq. (\ref{mm.1}), 
one obtains that the transmission matrix elements are given by

\beq
[ M ( E ; \chi ; \ell )]_{(j , \lambda ) ; (j' , \lambda' )} = \sum_{ \bar{j} = 1}^4 \: 
\sum_{ \rho = 1}^K [R ( E ; \chi ) ]_{(j , \lambda) ; (\bar{j} , \rho ) } 
[ L ( E ; \chi ) ]_{(\bar{j} , \rho ) ; ( j' , \lambda' ) } e^{ i \alpha_{ ( \bar{j} , \rho ) } \ell } 
\;\;\;\; . 
\label{mmq.1}
\eneq
\noindent
To compute ${\cal G} [ E ; \chi ; \ell ]$, we use the formula for the 
$M^B$-matrix in Eq. (\ref{mc.19}), which implies that ${\cal G} [ E ; \chi ; \ell  ]$
is given by the determinant of a $2N \times 2N$-matrix, whose entries are 
given by the matrix elements $[M ( E ; \chi : \ell ) ]_{(2k , \lambda ) ; (2k' , \lambda' )} $ with 
$k , k' = 1,2$. From Eq. (\ref{mmq.1}), we obtain 

\begin{eqnarray}
 [M ( E ; \chi : \ell ) ]_{(2k , \lambda ) ; (2k' , \lambda' )} &=& \sum_{\rho = 1}^K \biggl\{
 e^{ i \alpha_{ p , \rho} \ell} R_{(2k,\lambda) , ( 1 , \rho ) } 
 L_{ ( 1 , \rho ) , ( 2k' , \lambda' )} +  e^{ - i \alpha_{ p , \rho} \ell} R_{(2k,\lambda) , ( 2 , \rho ) } 
 L_{ ( 2 , \rho ) , ( 2k' , \lambda' )} \nonumber \\
 &+& e^{ - i \alpha_{ h , \rho} \ell} R_{(2k,\lambda) , ( 3 , \rho ) } 
 L_{ ( 3 , \rho ) , ( 2k' , \lambda' )} +  e^{  i \alpha_{ h , \rho} \ell} R_{(2k,\lambda) , 
 ( 4 , \rho ) }  L_{ ( 4 , \rho ) , ( 2k' , \lambda' )}  \biggr\}
 \:\:\:\: . 
 \label{mmaq.1}
\end{eqnarray}
\noindent
When computing the determinant of the matrix in Eq. (\ref{mmaq.1}), one readily
sees that it cannot contain a term proportional, for instance, to 
$e^{ 2 i \alpha_{ p , \rho} \ell}$, for any $\rho$. Indeed, in the determinant, 
a term of this form should arise from a sum of the form 
$\epsilon_{a_1 , \ldots , a_N}^{a_1' , \ldots , a_N'} e^{ 2 i\alpha_{ p ,\rho} \ell }
R_{ a_1 , ( 1 , \rho ) } L_{ (1,\rho) , a_1'} R_{ a_2 , ( 1 , \rho ) } L_{ (1,\rho) , a_2'} 
\ldots $, with $a_j$ and  $a_j'$ corresponding to a pair of indices such as 
$ ( j , \lambda ) $ and $(j' , \lambda' )$ and 
$\epsilon_{a_1 , \ldots , a_N}^{a_1' , \ldots , a_N'} $ being the 
fully antisymmetric tensor. Clearly, a term such as the one shown before
is equal to 0. Therefore, one obtains

\be
 {\cal G} [ E ; \chi ; \ell ] = \left[ \prod_{ \rho = 1}^K \sum_{ \{ a_\rho , b_\rho \}  \in 
\{ -1 ,0,  1 \} }\{   \delta_{a_\rho , 0 } \delta_{b_\rho , 0 } 
 + \delta_{| a_\rho | , 1 } \delta_{ | b_\rho | , 1 }  \}   
 \times
e^{ i [ a_\rho \alpha_p^{(\rho)} + 
b_{\rho } \alpha_h^{(\rho )} ] \ell } \right]
G_{\{  a_1, b_1 ;a_2, b_2 ; \ldots ;  a_K, b_K    \} } ( E ; \chi ) 
\;\;\;\; , 
\label{ft.3}
\ee
with  the coefficients $G_{\{  a_\rho , b_{\rho }    \} } ( E ; \chi )$
being fully determined by the $L$- and $R$-matrix 
elements. Note that, in writing Eq. (\ref{ft.3}), we have 
evidenced that the nonzero contributions are 
either characterized by $ a_\rho = b_{\rho }   = 0$, or by
$\{ a_\rho , b_{\rho } \}  \in 
\{ -1 , 1 \} $. From Eq. (\ref{ft.3}) we see that, in the specific case  
of  $K$ equivalent (that is, with the same dispersion relation) channels within C,
i.e., assuming that  $\alpha_{ j ,\rho}$ is  independent of $\rho$, 
the contributions to Eq. (\ref{ft.3}) can be grouped together,
so that one obtains 
\beq
{\cal G} [ E ; \chi ; \ell ] = \tilde{\sum}_{a,b=-K}^K G_{a,b} ( E ; \chi ) e^{ i [ a \alpha_p + 
b \alpha_h ] \ell} 
\:\:\:\: , 
\label{ss.1}
\eneq
\noindent
where $\tilde{\sum}$ means that the sum is taken over $ | a - b | = 0 $ (mod 2)
and the coefficients $G_{a,b} ( E ; \chi )$ being defined by comparing 
Eq. (\ref{ss.1}) to Eq. (\ref{ft.3}).
Eqs.(\ref{ft.3},\ref{ss.1}) are exact and provide the multi-channel generalization
of the  analogous  formulas of
Ref. [\onlinecite{giu_af}]. In  the following, we will use them to derive the 
 Josephson current in the long-junction limit. 
As the formal  manipulations required to recover 
the formulas for the Josephson current are, in general, different whether 
the $K$ channels are equivalent, or not, in the following we separately consider the 
case of equivalent and non-equivalent channels within C.

\subsection{The Josephson current in the case of equivalent channels within the central region}
\label{jos_equi}

In the case of $K$ equivalent channels, ${\cal G} [ E ; \chi ; \ell  ]$ is given in
Eq. (\ref{ss.1}).  In using Eq. (\ref{mc.24}) to compute $I  [ \chi ; T=0 ]$ in the long-junction
limit, we employ the same approximation used in Ref. [\onlinecite{giu_af}] in 
the single-channel case, that is, we 
set $\alpha_{p/h} \approx \alpha_F \pm i \omega / v$, with the Fermi momentum 
$\alpha_F = \sqrt{2 m \mu}$ and the Fermi velocity
$v = \alpha_F / m$, and  $m$ being the effective mass, $\mu$ the chemical potential. 
At the same time, we approximate  $G_{a,b} ( E ; \chi ) \approx G_{a,b} (E = 0 ; \chi ) 
\equiv G_{a,b} ( \chi )$, which is correct up to subleading contributions in 
$\ell^{-1}$ to the current. 
Eq. (\ref{mc.24}) eventually yields

\beq
I  [ \chi ; T=0 ] = \frac{2 e}{2 \pi} \: \int_{- \infty}^\infty \: d \omega \: 
\partial_\chi  \ln [ \sum_{ j = - K}^K P_{j+K} ( \chi ) e^{ - \frac{2 \omega \ell}{v}j }   ] 
+ \ldots 
\;\;\;\; , 
\label{ss.2}
\eneq
\noindent
with the coefficients $P_r ( \chi )$ fully determined by the matrix elements
of $ L ( E=0 ; \chi )$ and of $R ( E = 0 ; \chi  )$ and the ellipses corresponding to terms going 
to zero faster than  $\ell^{-1}$ in the large-$\ell$ limit, which we will neglect henceforth.  It can 
be shown that $P_0=P_{2K}=1$. (See Appendix D.) 
Switching to the integration variable 
$u = e^{ - \frac{2 \omega \ell}{v } }$, we obtain 

\beq
 I  [ \chi ; T=0 ] = \frac{2e}{2 \pi}\: \frac{v}{2 \ell} \; \int_{0}^\infty \; \frac{ d u}{
 u} \:  
 \left\{ \frac{ \sum_{j=1}^{2K-1} \partial_\chi P_j ( \chi ) u^j   }{
u^{2K} + \sum_{j=1}^{2K-1} P_j ( \chi ) u^j + 1 } \right\}
\:\:\:\: . 
\label{sol.11}
\eneq
\noindent
As discussed in detail in Appendix \ref{root_poly}, pertinently
computing the integral in Eq. (\ref{sol.11}) and observing that, denoting
with $u_j ( \chi )$ ($j=1 , \ldots , 2 K$) the roots of 
the equation ${\cal P} ( u ; \chi ) = u^{2K} + \sum_{j=1}^{2K-1} P_j ( \chi ) u^j + 1 = 0$, one
obtains $\prod_{j=1}^{2K} u_j ( \chi ) = 1$, one eventually gets

\beq
 I  [ \chi ; T=0 ] =  \frac{ e v}{4 \pi \ell} \: \sum_{j=1}^{2K} \partial_\chi 
 \ln^2 [ u_j ( \chi ) ] 
 \:\:\:\: . 
 \label{sol.12}
 \eneq
 \noindent
 While, in general, the coefficients  $G_{a,b} ( E ; \chi )$ are complicated
functions of the $L (E ; \chi)$- and of the  $R(E ; \chi)$-matrix elements, Eq. (\ref{sol.12})
only involves quantities evaluated at the Fermi level. This allows for 
building a simplified algorithm for constructing the polynomial 
${\cal P} ( u ; \chi  ) = u^{2K} + \sum_{j=1}^{2K-1} P_j ( \chi ) u^j + 1 $,
which we discuss in detail in Appendix \ref{co_put}. 
 
The generalization of  Eq. (\ref{sol.12}) to $T$ finite, but still 
much lower than the superconducting gap, can be again 
worked out by  deforming  the integration path in the complex energy plane, so that the final 
integral over the imaginary axis is traded for a sum of integrals over small 
circles surrounding the points $i \omega_\nu$ over the imaginary axis (see Fig. \ref{fig_multi1}), with 
  $\omega_\nu  $ being the 
$\nu^{\rm th}$ fermionic Matsubara frequency, $\omega_\nu = \pi T ( 2 \nu + 1)$.\cite{giu_af,ishi,been.0,furus}
To work out the modification of Eq. (\ref{ss.2}) at finite-$T$, we also consider that, 
due to the particle-hole symmetry of the Bogoliubov- de Gennes equations near by the
Fermi level, one obtains that $P_{K+j} ( \chi ) = P_{K-j} ( \chi )$. Thus, performing the integrals
over each circle and adding up the results, one obtains  

\beq
I [ \chi ; T ] = 2 e T \: \sum_{ \nu = - \infty}^\infty \: \partial_\chi \left\{ \ln \left[ 
2 \cosh \left( \frac{2 K \omega_\nu \ell}{v} \right) + 2 \sum_{ j = 0}^{K-1} P_j ( \chi ) 
\cosh \left( \frac{2 j \omega_\nu \ell}{v}\right)  \right] \right\} + \ldots 
\:\:\:\: , 
\label{ft.1}
\eneq
\noindent
with, again, the ellipses corresponding to terms going 
to zero faster than  $\ell^{-1}$ in the large-$\ell$ limit.
As $\ell T / v \to 0$, the sum in Eq. (\ref{ft.1}) can be traded for an
integral over a continuous variable $\omega$, thus leading back to Eq. (\ref{ss.2}).
For $N=K=1$,  it is easy to check that one obtains 
Eq. (53) of [\onlinecite{giu_af}] for the finite-temperature dc Josephson current 
in this case. Finally, in the regime  $\ell T / v \gg 0$, $I [ \chi ; T ]$
exhibits an exponential decay in $T$, again consistent with the result of 
[\onlinecite{giu_af}].

\begin{figure}
\includegraphics*[width=0.75\linewidth]{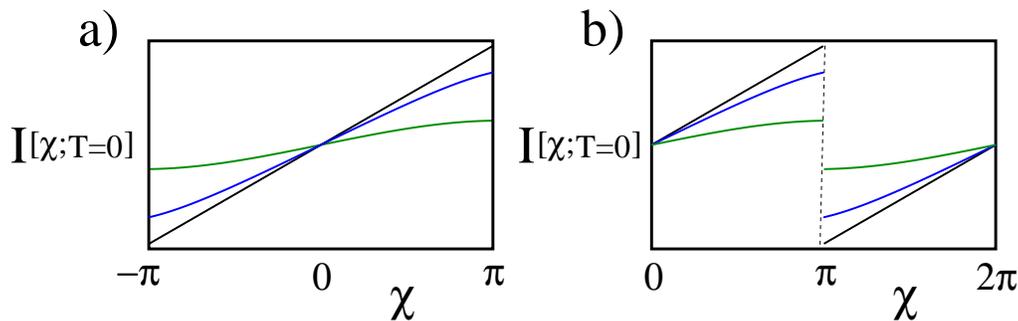}
\caption{{\bf a):} Plot of $I  [ \chi ; \theta ;  T=0 ]$ {\it vs.} $\chi$, as in Eq. (\ref{dc.1.10}), 
for $- \pi \leq \chi \leq \pi$, 
with $\ell$ fixed, $\alpha_F \ell = .43 \pi$ and, respectively, $\theta = 0$ (black curve), 
$\theta = 0.2 \pi$ (blue curve), $\theta = 0.4 \pi$ (green curve);  
{\bf b):} Same as in panel {\bf a)}, but for $0 \leq \chi \leq 2 \pi$, to evidence the 
finite discontinuity in $I  [ \chi ; \theta ; T=0 ] $ at $\chi = \pi$.  } \label{fig_multi2}
\end{figure}

From Eqs.(\ref{sol.12},\ref{ft.1}), we see that the key quantity needed to 
compute the current, both at $T=0$ and at $T>0$, is the polynomial 
${\cal P} ( u ; \chi ) $. In Appendix \ref{co_put} we discuss in detail the algorithm
for constructing ${\cal P} ( u ; \chi )$ in general and carry out the whole
calculation in the specific case $N=1$. In particular, we show that, provided $K \geq 2$, 
the calculation can be always reduced to  a model with $K=2$ channels within 
C coupled to the L- and to the R- channel with 
strengths of the form $([ t_L ]_1 ,[ t_L ]_2 ) = t_L ( \cos ( \theta ) , 
\sin ( \theta  ) )$ and $([ t_R ]_1 ,[ t_R ]_2 ) = t_R ( 1 , 
0 )$, respectively. As a specific model calculation,  we 
explicitly compute Eq. (\ref{sol.12}) in the case 
in which the two channels within C effectively coupled to the leads
both exhibit perfect Andreev reflection. In this case, one obtains 
 (see Appendix \ref{co_put} for details)

\beq
{\cal P}_\theta ( u ; \chi  ) = u^4 + 1 - 2 \cos^2 ( \theta ) [ \cos ( 2 \alpha_F \ell ) +\cos ( \chi ) ] ( u^3 + u ) 
+ 2 \{ \cos ( 2 \theta ) + 2 \cos^2 ( \theta ) \cos ( 2 \alpha_F \ell ) \cos ( \chi ) \} u^2  
\:\:\:\: , 
\label{dc.1.6}
\eneq
\noindent
with the suffix $_\theta$ added to evidence the dependence of
${\cal P }$ on this parameter, as well. 
The equation ${\cal P}_\theta ( u ; \chi  ) = 0$  may be straightforwardly solved by means of elementary
algebraic techniques. Its roots are given by

\begin{eqnarray}
 u_1 (\theta ;  \chi )  &=& z_1(  \theta ;  \chi ) + i \sqrt{1 - z_1^2(  \theta ;  \chi )} \nonumber \\
u_2 ( \theta ;  \chi )  &=& z_1 (  \theta ;  \chi )- i \sqrt{1 - z_1^2(  \theta ;  \chi )} \nonumber \\
 u_3 ( \theta ;  \chi )  &=& z_2  ( \theta ;  \chi )+ i \sqrt{1 - z_2^2(  \theta ;  \chi )} \nonumber \\
u_4 ( \theta ;  \chi )  &=& z_2 (  \theta ;  \chi )- i \sqrt{1 - z_2^2(   \theta ;  \chi )}
\:\:\:\: , 
\label{dc.1.7} 
\end{eqnarray}
\noindent
with

\begin{eqnarray}
z_1 (  \theta ;  \chi )  &=&  \frac{1}{2} \left\{  \cos^2 ( \theta ) [ \cos ( 2 \alpha_F \ell ) +\cos ( \chi ) ]
+ \sqrt{ 4 \sin^2 ( \theta ) - 4 \cos^2 ( \theta ) \cos ( 2 \alpha_F \ell ) \cos ( \chi ) 
+ \cos^4 ( \theta ) [ \cos ( 2 \alpha_F \ell ) + \cos ( \chi ) ]^2} \right\} \nonumber \\
z_2 (  \theta ;  \chi )  &=&  \frac{1}{2} \left\{  \cos^2 ( \theta ) [ \cos ( 2 \alpha_F \ell ) + \cos ( \chi ) ]
- \sqrt{ 4 \sin^2 ( \theta ) - 4 \cos^2 ( \theta ) \cos ( 2 \alpha_F \ell ) \cos ( \chi ) 
+ \cos^4 ( \theta ) [ \cos ( 2 \alpha_F \ell ) + \cos ( \chi ) ]^2} \right\} \nonumber \\
\:\:\:\: . 
\label{dc.1.a} 
\end{eqnarray}
\noindent
From Eqs.(\ref{dc.1.7}) we see that it is  possible to write 

\beq
u_{1,2} ( \theta , \chi ) = e^{ \pm i \vartheta_1 [   \theta ; \chi ] } \;\;\; , \;\;
u_{3,4} ( \theta , \chi ) = e^{ \pm i \vartheta_2 [   \theta ; \chi ] }
\;\;\;\; , 
\label{dc.1.8}
\eneq
\noindent
with $\vartheta_j [   \theta ; \chi ]  = {\rm arccos} \{  z_j (  \theta ;  \chi ) \}$. 
Thus, we eventually obtain  that the dc Josephson current
is given by (making explicit the dependence on the parameter $\theta$, as well)

\beq
I  [ \chi ; \theta  ; T=0 ] = - \frac{  e v}{\pi \ell} \partial_\chi \{ \vartheta^2_1 [ \theta ; \chi ] + 
\vartheta_2^2 [ \theta ; \chi ] \}
\:\:\:\:.
\label{dc.1.10}
\eneq
To check the consistency of Eq. (\ref{dc.1.10}), we   notice that, 
for $\theta = 0$,  $I  [ \chi ; \theta  = 0 ; T=0 ]$  reduces back to Ishii's sawtooth behavior
\cite{ishi} corresponding to perfect Andreev reflection at both boundaries. At  variance, 
for $\theta = \frac{\pi}{2}$ one obtains $I  [ \chi ;\theta  = \frac{\pi}{2} ; T=0 ] = 0$, as it is appropriate 
to a situation where only channel-1 within C is coupled to the left-hand lead and
only channel-2 is coupled to the right-hand lead. To evidence the effect of a  finite value of $\theta$
such that $0< \theta < \frac{\pi}{2}$, in Fig. \ref{fig_multi2} we plot $I  [ \chi ;\theta ; T=0 ]$ {\it vs.} $\chi$ 
for three values of $\theta$, including $\theta = 0$ (see caption for details). It is
interesting to note that a finite discontinuity takes place at $\chi = \pi$ (mod $2 \pi$) for
any value of $\theta$ and that   $I [ \chi = \pi^- ; \theta ; T=0 ] - 
I [ \chi = \pi^+ ; \theta ; T=0 ] 
\propto \cos^2 ( \theta)$. This is a typical feature of  junctions exhibiting 
perfect Andreev reflection at the SN interfaces; formally, it is a consequence of the fact that, as it can be 
readily seen from Eqs.(\ref{dc.1.a}),  $ z_2 (  \theta ; \chi ) $ always reaches the value $-1$ as 
$\chi  \to \pi $ (mod $2 \pi$), irrespectively of the values of $\alpha_F  \ell $ and $\theta$. 
Thus, though  $\vartheta_2 [ \theta ; \chi ]$ is continuous at $\chi = \pi$, it exhibits a cusp, 
with a corresponding finite discontinuity in its derivative. This is what determines 
the discontinuity in the plots of $I  [ \chi ; T=0 ]$ {\it vs.} $\chi$ in Fig. \ref{fig_multi2}.
An important remark about Eq. (\ref{dc.1.10}) is that, though, at a first glance, 
it looks similar to what one would get by only summing the contributions to 
$I [ \chi ;\theta ; T = 0 ]$ arising from ABS's near  the Fermi energy, 
in fact, as a result of the cancellations between large contributions to 
the current from states far from the Fermi energy, the result is 
exact, to leading order in $\ell^{-1}$, as we proved before. 

  While in ``conventional'' multi-channel junctions the various channels
do not exhibit equivalence, as they
 typically have different Fermi velocities, an SNS junction with two equivalent 
channels can be realized for instance by connecting a non-chiral metallic carbon nanotube 
to two spinful $s$-wave superconductors. Electrons around the two non-equivalent 
Dirac points in the single-electron spectrum of the carbon nanotube\cite{nontube} 
act as two spinful independent channels, thus realizing the system we discuss 
in detail in appendix \ref{co_put} and in which, in this section, we explicitly solve for
$ I [ \chi ; T = 0  ]$ in the special case of pure Andreev reflection in each channel
coupled to the superconducting leads.

\noindent

When there is no equivalence between the channels
within C,  from the discussion we make in Section \ref{multi_cha},
one expects  that the current in the long-junction limit is still determined by reflection 
coefficients at the Fermi level, even in the case of inequivalent 
channels. However, as we are going to outline in the following section, it is in 
general not possible to resort to a simple and compact analytical expression, such 
as the one in Eqs.(\ref{sol.11},\ref{ft.1}) and, therefore, one has to numerically evaluate the 
resulting integral which is expected to depend on a number of parameters, including 
the asymmetries between the channels.

\subsection{The Josephson current in the case of inequivalent channels within the central region}
\label{long_junctioni}

In the case of inequivalent channels within C, ${\cal G} [ E ; \chi ; \ell ]$ 
is given in Eq. (\ref{ft.3}). 
In the large $\ell$ limit one may again perform the approximation
used in Subsection \ref{jos_equi} and discussed in  [\onlinecite{giu_af}]. As a 
result, Eq. (\ref{mc.24}) for $I  [ \chi ; T=0 ]$ generalizes to

\begin{eqnarray}
I  [ \chi ; T=0 ]   &=&  
\frac{2e}{2 \pi} \: \frac{U}{ \ell}\: \int \: d z \: \partial_\chi \{ \ln [ \prod_{ \rho = 1}^K
\sum_{ \{ a_\rho , b_\rho \} \in \{ - 1 , 0 , 1 \} } [ [   \delta_{a_\rho , 0 } \delta_{b_\rho , 0 } + 
 \delta_{ | a_\rho | , 1 } \delta_{| b_\rho | , 1 } ]  
e^{ i ( a_\rho - b_\rho )  \alpha_F^{(\rho)}  \ell } e^{ - w_\rho  ( a_\rho + b_\rho ) z} ] 
\nonumber \\
 & \times&   \bar{G}_{\{ a_1 , b_1 , \ldots , a_K , b_K \} } (  \chi )   \}
 + \ldots  
 \;\;\;\; , 
 \label{ft.4}
 \end{eqnarray}
 \noindent
with $\bar{G}_{\{ a_1  , b_1 , \ldots , a_K , b_K  \} } (  \chi ) = G_{\{ a_1 , b_1 , \ldots , 
a_K , b_K  \} }  ( E = 0 ; \chi )$,  $\alpha_F^{(\rho)} , v^{(\rho)}$ being respectively the Fermi momentum and 
the Fermi velocity for channel-$\rho$, $U^K = \prod_{\rho = 1}^K v^{(\rho)}$, and 
$w_\rho = U/ v^{(\rho)}$. Note that, in Eq. (\ref{ft.4}), we have introduced the rescaled
integration variable $z = \omega \ell / U$ and that, as in the similar equations above, the ellipses correspond to 
subleading contributions going to zero faster than $\ell^{-1}$ in the large-$\ell$ limit. 
Similarly,  Eq. (\ref{ft.1}) for $I  [ \chi ; T ]$ 
now generalizes to

\begin{eqnarray}
I  [ \chi ; T ] &=& 2 e T \: \sum_{ \nu = - \infty}^\infty  \:  \partial_\chi \{ \ln [ \prod_{ \rho = 1}^K
\sum_{ \{ a_\rho , b_\rho \} \in \{ - 1 , 0 , 1 \} } [ [   \delta_{a_\rho , 0 } \delta_{b_\rho , 0 } + 
 \delta_{ | a_\rho | , 1 } \delta_{| b_\rho | , 1 } ]  
e^{ i ( a_\rho - b_\rho )  \alpha_F^{(\rho)}  \ell } e^{ - w_\rho  ( a_\rho + b_\rho )  \frac{ 
 \omega_\nu \ell}{U}  }   ] 
\nonumber \\
 & \times&   \bar{G}_{\{ a_1 , b_1 , \ldots , a_K , b_K \} } (  \chi )   \}
 \;\;\;\; . 
 \label{ft.4bis}
 \end{eqnarray}
 \noindent
As it clearly appears from Eqs.(\ref{ft.4},\ref{ft.4bis}), 
the general result that in the long-junction limit the current only
depends on backscattering amplitudes at the Fermi level holds in
the case of inequivalent channels, as well. The key function 
one has to derive, in order to compute $ I  [\chi ; T ]$ in the 
long-junction limit, is the function $\Phi [ \omega ; \chi  ] $,
defined as 

\begin{eqnarray}
\Phi [ \omega ; \chi  ] &=& 
\prod_{ \rho = 1}^K \{
\sum_{ \{ a_\rho , b_\rho \} \in \{ - 1 , 0 , 1 \} } [ [   \delta_{a_\rho , 0 } \delta_{b_\rho , 0 } + 
 \delta_{ | a_\rho | , 1 } \delta_{| b_\rho | , 1 } ]  e^{ i ( a_\rho - b_\rho )  \alpha_F^{(\rho)}  \ell } 
e^{ - w_\rho  ( a_\rho + b_\rho ) \omega} ] 
\nonumber \\&\times& 
 \bar{G}_{\{ a_1 , b_1 , \ldots , a_K , b_K \} } (  \chi ) ]  
\}
\:\:\:\: , 
\label{newy.1}
\end{eqnarray}
\noindent
with the coefficients $\bar{G}_{\{ a_1 , b_1 , \ldots , a_K , b_K \} } (  \chi )$
defined as in Eq. (\ref{ft.4}). In Appendix \ref{phi_omchi} 
we discuss the systematic procedure to construct $\Phi [ \omega ; \chi ]$:
clearly, the final result will apply in general, including the case
of equivalent channels within C. In this latter case, however, as we 
discuss in subSection  \ref{long_junctioni}, once expressed in 
terms of the variable $u = e^{  - 2 \omega }$, $\Phi [ \omega ; \chi  ]$
reduces to the $2K$-degree polynomial ${\cal P} ( u ; \chi )$ in the 
 variable $u$.
 
\begin{figure}
\includegraphics*[width=.55\linewidth]{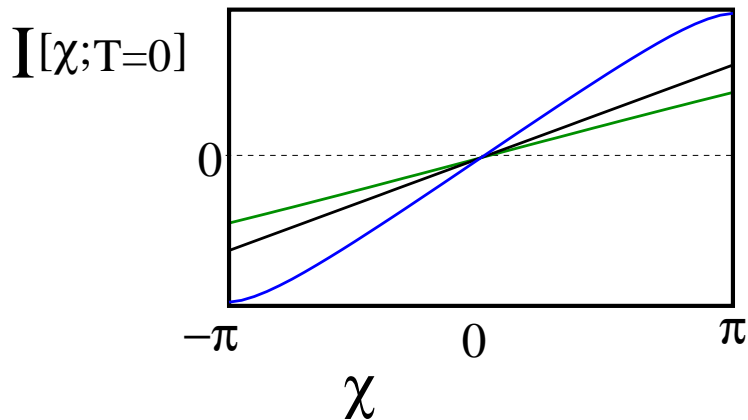}
\caption{Plot of $I  [ \chi ;  T=0 ]$ {\it vs.} $\chi$, numerically
computed as from Eq. (\ref{ft.4}) with $\Phi [ \omega ; \chi  ]$ given in 
Eq(\ref{ineq.1}). The parameters are chosen so that $\alpha_F^{(1)} \ell =
.48 \pi, \alpha_F^{(2)} \ell =
.51 \pi, \varphi = \pi / 10$, while $w = .5$ for the blue curve, $w=1$
for the black curve, $w=1.5 $ for the green curve.} \label{fig_multi3}
\end{figure}

As an example of
the effectiveness of our procedure,  we compute 
$I  [ \chi ; T=0 ]$ for $N=1, K=2$ in the case in which the two
channels within C are characterized by Fermi momenta $\alpha_F^{(1)} , 
\alpha_F^{(2)}$ and by Fermi velocities $v^{(1)}, v^{(2)}$,
respectively, and the couplings at the SN interfaces are 
$( [t_L]_1 , [ t_L]_2 ) = t_L ( \cos ( \varphi ) , \sin ( \varphi )) $ and
$( [t_R]_1 , [ t_R]_2 ) = t_R ( \cos ( \varphi ) , \sin ( \varphi )) $.
As we are going to show in the following, the absence of the 
symmetry between the two channels makes even this simple case
quite interesting to consider. To derive $\Phi [ \omega ; \chi ]$, 
we use the formula in Eq. (\ref{ft.8}) of Appendix  \ref{phi_omchi} 
which, in the specific case we are dealing with, yields 

\begin{eqnarray}
 \Phi [ \omega ; \chi   ] &=& 4 \cos^4 ( \varphi) \cos ( 2 \alpha_F^{(1)} \ell )
 \left[ \cos ( \chi ) - \cosh ( 2 \omega w ) \right] + 
 4 \sin^4 ( \varphi) \cos ( 2 \alpha_F^{(2)} \ell )
 \left[ \cos ( \chi ) - \cosh \left( \frac{ 2 \omega }{ w} \right) \right]
 \nonumber \\
 &-& 4 \left[ \cos^4 ( \varphi ) \cosh \left( \frac{2 \omega}{w} \right) 
 + \sin^4 ( \varphi ) \cosh ( 2 \omega w ) \right] \cos ( \chi ) + 
 2 \cosh \left( 2 \omega w + \frac{2 \omega}{w} \right) 
 + [ 1 + \cos ( 4 \varphi ) ] 
 \cosh \left( 2 \omega w - \frac{2 \omega}{w} \right) \nonumber \\
 &+& 8 \sin^2 ( 2 \varphi ) \left\{ \cosh \left( \frac{\omega}{w} \right) 
 \cosh ( \omega w ) \sin ( \alpha_F^{(1)} \ell )\sin ( \alpha_F^{(2)} \ell )
 \sin^2 \left( \frac{\chi}{2} \right) -  \sinh \left( \frac{\omega}{w} \right) 
 \sinh ( \omega w ) \cos ( \alpha_F^{(1)}\ell  )\cos ( \alpha_F^{(2)} \ell )
 \cos^2 \left( \frac{\chi}{2} \right) \right\} \nonumber \\
 \:\:\:\: , 
 \label{ineq.1}
\end{eqnarray}
\noindent
with $w = \sqrt{\frac{v^{(2)}}{v^{(1)}}}$. Clearly, the ``relative contribution''
of the two channels within C to the total current depends on the angle $\varphi$. 
For instance, if $\varphi$ is closer to 0 than to $\frac{\pi}{2}$, channel-1 is
expected to provide a contribution higher than the one provided by channel-2. Thus, 
on tuning the asymmetry between the two channels, in this case we expect the current to 
increase (decrease), if the asymmetry ``weights'' more the contribution from
channel-1 (channel-2). To check this point, in Fig. \ref{fig_multi3}, 
we plot the current at fixed $\varphi$, $I  [ \chi ; T=0 ]$ {\it vs.} $\chi$, numerically computed  
using the formula for $\Phi [ \omega ; \chi ]$ in Eq. (\ref{ineq.1}), with   
$\varphi = \pi / 10$ and the other parameters 
fixed as detailed in the caption of the figure. As expected,  
at fixed $\chi$, we see that the smaller is $w$ (that is, the higher is 
the Fermi velocity in channel-1 with respect to the one in 
channel-2), the higher is the current. 

\noindent
As we showed, once $\Phi [ \omega ; \chi ]$ is computed as discussed 
in Appendix \ref{phi_omchi}, Eqs.(\ref{ft.4},\ref{ft.4bis}) 
provide a simple and effective tool to  
compute $I  [ \chi ; T=0 ]$ and  $I [ \chi ; T ]$ for generic 
values of the parameters. The important information they encode is 
that, in the long junction limit, the current is fully expressed  only in terms of 
reflection amplitudes computed at the Fermi energy. Once the reflection amplitudes are 
known, the integral and/or the sum can be computed numerically, which (especially for 
a long junction) is enormously  simpler 
than performing a sum over contributions from all kind of states at any
energy.\cite{perfetto} Our result holds in general, 
independently of the symmetry between the channels within C and, in 
the symmetric case, it is possible to work out simple closed-form formulas for 
the current, as the ones we provide in Eqs.(\ref{sol.12},\ref{ft.1}).

 \section{Conclusions}
 \label{concl}
 
In this paper, we go through a systematic application of the analytic properties of 
the S-matrix for a long multi-channel SNS junction, to show that the dc Josephson current 
across the junction at low temperatures can be fully expressed in terms of scattering amplitudes at the 
Fermi level only. When the dispersion relations for the channels within the 
central region are equal to each other, the current can be expressed in terms of  a simple, closed-form formula,
given in Eq. (\ref{sol.12}) in the zero-temperature limit, in Eq. (\ref{ft.1}) at finite temperature. 
In general, the current can still be simply computed, by evaluating integrals involving 
only scattering amplitudes at the Fermi level. Besides providing a simple and effective algorithm for 
computing the current, our results   justify resorting to a  low energy 
Hamiltonian approach,\cite{ACZ} which is crucial for treating 
Luttinger liquid interaction effects. While we choose a model Hamiltonian in which the leads 
are pictured as one-dimensional s-wave superconductors, our results are expected to readily
generalize to situations in which the leads are realized, for instance, as topological p-wave
superconductors,  where the dc Josephson current is, in general, strongly affected by 
the possible presence of emerging Majorana fermions at the SN-interfaces.
 
We would like to thank A. Nava and P. Lucignano for helpful discussions. 
DG would like to thank the Department of Physics and Astronomy of the
University of British Columbia for the kind hospitality at various stages
of this work. This research was supported in part by NSERC and CIfAR.

\appendix

\section{Scattering solutions of the Bogoliubov - de Gennes equations }
\label{models}

In this appendix we review the derivation of the single-quasiparticle and
of the single-quasihole scattering solutions to the Bogoliubov - de Gennes
equations for a multi-channel SNS junction. In particular, we first 
derive the asymptotic form of the scattering solutions within the 
superconducting leads, which is expected to apply to a generic junction, without
specializing to the long-junction limit. Thereafter, we will discuss in detail the 
case of a long SNS junction.

\subsection{Asymptotic solutions within the superconducting leads}
\label{models_details}

We perform our derivation  within a straightforward
multi-channel generalization of the   continuum one-dimensional model for a 
spinful superconductor discussed in [\onlinecite{btk}]. Besides 
the simplifying assumption $N_R = N_L \equiv N$, since  by a phase redefinition we can always choose the phases of 
the order parameter to be equal and opposite in the left/right leads, we also require 
 that the phase  difference $\chi$ between the leads is uniformly distributed between the 
two sides, namely,  that the phase of the superconducting order parameter
 $ \to \frac{\chi}{2}$ for $x \to -  \infty$ and 
$\to - \frac{\chi}{2}$ for $x \to \infty$, and that the 
superconducting gap $\Delta$ is the same for all the channels. Thus, 
the second-quantized Hamiltonians for leads L and R, 
$H_L$ and $H_R$, are respectively given by

\begin{eqnarray}
 H_L-\mu {\cal  N}_L &=& \int_{ x \in L}  \: d x \:  \sum_{ \lambda = 1}^N  \: \left\{ \sum_\sigma
 \Psi_{L , \lambda , \sigma}^\dagger 
 ( x ) h_{0 , \lambda} ( x )  \Psi_{L , \lambda , \sigma}  ( x )
 + \Delta e^{  \frac{i}{2} \chi }  \Psi_{L , \lambda , \uparrow}  ( x ) 
 \Psi_{L , \lambda , \downarrow}  ( x ) + 
 \Delta e^{ -  \frac{i}{2} \chi }  \Psi_{L , \lambda ,\downarrow }^\dagger   ( x ) 
 \Psi_{L , \lambda , \uparrow }^\dagger   ( x ) \right\} \nonumber \\
  H_R -\mu {\cal N}_R &=& \int_{ x \in R}  \: d x \:  \sum_{ \lambda = 1}^N \: \left\{ \sum_\sigma 
 \Psi_{R , \lambda , \sigma}^\dagger 
 ( x ) h_{0 , \lambda} ( x )  \Psi_{R , \lambda , \sigma}  ( x )
 + \Delta e^{  - \frac{i}{2} \chi }  \Psi_{R , \lambda , \uparrow}  ( x ) 
 \Psi_{R , \lambda , \downarrow}  ( x ) + 
 \Delta e^{   \frac{i}{2} \chi }  \Psi_{R , \lambda ,\downarrow }^\dagger   ( x ) 
 \Psi_{R , \lambda , \uparrow }^\dagger   ( x ) \right\}
 \;\;\;\; , 
 \label{newvel.01}
\end{eqnarray}
\noindent
with $ \Psi_{L , \lambda , \sigma}  ( x ) ,  \Psi_{R , \lambda , \sigma}  ( x )$
being the fermion annihilation operators for an electron in channel-$\lambda$ 
with spin $\sigma$ in lead L and R, respectively. 
$h_{0 , \lambda} ( x ) = - \frac{1}{2 m_{S , \lambda}} \frac{d^2}{d x^2 } +
V_{S , \lambda}$ is the 
normal lead Hamiltonian in channel-$\lambda$, $m_{S , \lambda}$ and 
$V_{S , \lambda}$ are the corresponding effective electron mass
and  potential, respectively. The BDG
equations within L and  R   are derived starting from the 
Bogoliubov-Valatin transformations, which enable us to express 
an energy eigenmode operator of $H_L$, $\gamma_{ L, \lambda , \sigma} ( E ) $, as 

\beq
\gamma_{  L , \lambda ,  \sigma} ( E ) = \int_{x \in L} \: d x \: \{ u_{L , \lambda , E}  ( x ) 
\Psi_{ L , \lambda , \sigma} ( x ) + \sigma v_{L , \lambda , E} ( x ) 
\Psi_{ L , \lambda , \sigma}^\dagger  ( x ) \}
\;\;\;\; . 
\label{novelk.1}
\eneq
\noindent
Requiring that $ [ \gamma_{  L ,  \lambda , \sigma} ( E ) , H_L ] =
E  \gamma_{  L ,  \lambda , \sigma} ( E ) $
yields the BDG equations for the wavefunctions within L, $u_{L , \lambda , E} ( x ) 
, v_{L , \lambda , E} ( x )$, given by

\begin{eqnarray}
 h_{0 , \lambda} ( x ) u_{L , \lambda , E }  ( x ) + \Delta e^{\frac{i}{2}  \chi}  v_{L , \lambda , E}
 ( x ) &=&  E u_{L , \lambda , E }  ( x ) \nonumber \\
 \Delta e^{ - \frac{i}{2}  \chi}  u_{L , \lambda , E}  ( x ) - h_{0 , \lambda} ( x ) 
 v_{L , \lambda , E}  ( x )  &=& E v_{L , \lambda , E}( x ) 
 \:\:\:\: , 
 \label{a1.1}
\end{eqnarray}
\noindent
with $x \in L$. 
Similarly, writing an energy eigemode operator of $H_R$, $\gamma_{ R, \lambda , \sigma} ( E ) $, as 

\beq
\gamma_{  R , \lambda ,  \sigma} ( E ) = \int_{ x \in R} \: d x \: \{ u_{R , \lambda , E} ( x ) 
\Psi_{ R , \lambda , \sigma} ( x ) + \sigma v_{R , \lambda , E} ( x ) 
\Psi_{ R , \lambda , \sigma}^\dagger  ( x ) \}
\;\;\;\; , 
\label{novelk.2}
\eneq
\noindent
  and requiring that $ [ \gamma_{  R , \lambda ,  \sigma} ( E ) , H_R ] = 
  E  \gamma_{  R , \lambda ,  \sigma} 
  ( E ) $ yields the BDG equations for the wavefunctions within R, $u_{R , \lambda , E} ( x ) 
, v_{R , \lambda , E} ( x )$, given by

\begin{eqnarray}
 h_{0 , \lambda} ( x ) u_{R , \lambda , E }  ( x ) + \Delta e^{- \frac{i}{2}  \chi}  v_{R , \lambda , E}
 ( x ) &=&  E u_{R , \lambda , E }  ( x ) \nonumber \\
 \Delta e^{  \frac{i}{2}  \chi}  u_{R , \lambda , E }  ( x ) - h_{0 , \lambda} ( x ) 
 v_{R , \lambda  , E }  ( x )  &=& E v_{R , \lambda , E}( x ) 
 \:\:\:\: , 
 \label{bb1.1}
\end{eqnarray}
\noindent
with $x \in R$. Thus, one sees that a scattering solutions at energy  $E$ asymptotically obeys Eqs.(\ref{a1.1})
within L and Eqs.(\ref{bb1.1}) within R. As a consequence,  for each channel $\lambda$ 
one  finds four independent solutions to Eqs.(\ref{a1.1}): a  forward/backward particle-like- ($_{1,2}$) 
 and a forward/backward hole-like ($_{3,4}$) solution, respectively given by

 \beq
 \left[ \begin{array}{c} u_{L , \lambda , E}  ( x ) \\ v_{L , \lambda , E} ( x ) \end{array} \right]_{1,2} 
 = 
 \left[ \begin{array}{c} \cos \left( \frac{\Psi}{2} \right) \\ 
          - e^{\frac{i}{2}  \chi} \sin \left( \frac{\Psi}{2} \right) 
        \end{array} \right] e^{ \pm i \beta_{p , \lambda} x } \;\;\; , \;\;
        \left[ \begin{array}{c} u_{L , \lambda, E} ( x ) \\ v_{L , \lambda, E} ( x ) \end{array} \right]_{3,4} 
=
 \left[ \begin{array}{c} - e^{-\frac{i}{2}  \chi} \sin \left( \frac{\Psi}{2} \right)  \\
 \cos \left( \frac{\Psi}{2} \right) \\          
        \end{array} \right] e^{ \mp i \beta_{h , \lambda} x }     
\:\:\:\: , 
\label{sol.1}
\eneq
\noindent
with $\beta_{p/h , \lambda}^2 = 2 m_{S , \lambda} \{ \mu_{S , \lambda}  \pm ( E^2 -  \Delta^2 )^\frac{1}{2} \}$ 
and $\Psi \equiv -\arcsin (\Delta /E)$.  
Similarly, for each channel one   finds four analogous independent solutions to Eqs.(\ref{bb1.1}), 
given by

 \beq 
 \left[ \begin{array}{c} u_{R , \lambda, E} ( x ) \\ v_{R , \lambda, E} ( x ) \end{array} \right]_{1,2} 
= 
 \left[ \begin{array}{c} \cos \left( \frac{\Psi}{2} \right) \\ 
          - e^{-\frac{i}{2}    \chi} \sin \left( \frac{\Psi}{2} \right) 
        \end{array} \right] e^{ \pm i \beta_{p , \lambda} ( x - \ell) }    
  \;\;\; , \;\;
  \left[ \begin{array}{c} u_{R , \lambda, E} ( x ) \\ v_{R , \lambda, E} ( x ) \end{array} \right]_{3,4} =
 \left[ \begin{array}{c} - e^{  \frac{i}{2}   \chi} \sin \left( \frac{\Psi}{2} \right)  \\
 \cos \left( \frac{\Psi}{2} \right)           
        \end{array} \right] e^{ \mp i \beta_{h , \lambda}  ( x - \ell) }        
\:\:\:\: , 
\label{sol.2}
\eneq
\noindent
where the right lead is at $x>\ell$. 
Thus, in each channel $\lambda$,  a generic wavefunction within L (R), 
 $ \left[ \begin{array}{c} u_{L(R),\lambda, E} ( x ) \\ 
                                  v_{L(R),\lambda, E} ( x ) 
                                 \end{array} \right]$, can be written as 
a linear superpositions of the four kinds of plane  wave quasiparticle and 
quasihole solutions in Eqs.(\ref{sol.1},\ref{sol.2}) so that, in general, 
 one obtains

\beq
\left[ \begin{array}{c} u_{L , \lambda, E} ( x ) \\ 
        v_{L , \lambda, E} ( x ) 
       \end{array} \right] = \sum_{ j = 1}^4 A_{j, \lambda}^- (E;\chi)  \left[ \begin{array}{c} u_{L , \lambda, E} ( x ) \\ 
        v_{L , \lambda, E} ( x ) 
       \end{array} \right]_j \;\;\; , \;\; (x \in L)
       \;\;\;\; , 
       \label{mc.4}
       \eneq
       \noindent
and 
\beq
\left[ \begin{array}{c} u_{R , \lambda, E} ( x ) \\ 
        v_{R , \lambda, E} ( x ) 
       \end{array} \right] = \sum_{ j = 1}^4 A_{j, \lambda}^+ (E;\chi) \left[ \begin{array}{c} u_{R , \lambda, E} ( x ) \\ 
        v_{R , \lambda, E} ( x ) 
       \end{array} \right]_j \;\;\; , \;\; (x \in R)
       \;\;\;\; . 
       \label{mc.5}
       \eneq
       \noindent 
The transmission matrix $M ( E ; \chi )$ relates the $A_{j , \lambda}^+(E;\chi) $-amplitudes to the $A_{j , \lambda}^-(E;\chi) $-ones.
Thus, it appears natural to label the $M$-matrix elements with two pairs of
indices, $( j , \lambda) , ( j' , \lambda' )$, referring to the 
quasiparticle character and to the channel, respectively, so that 
the  matrix elements  $[ M ( E ; \chi )]_{(j,\lambda) , ( j' , \lambda' ) }$ satisfy

\beq
A_{j , \lambda}^+ ( E ; \chi ) = 
\sum_{j' = 1}^4 \sum_{\lambda' = 1}^N \: [M ( E ; \chi ) ]_{(j,\lambda) , ( j' , \lambda' )} 
A_{j' , \lambda'}^- ( E ; \chi ) 
\:\:\:\: . 
\label{mc.6}
\eneq
\noindent
At variance, the $S$-matrix relates to each other incoming ($in$) and outgoing ($out$) 
quasiparticle amplitudes. These are related to the $A_{j , \lambda}^\pm ( E ; \chi )$-amplitudes 
as 

\beq
\left[\begin{array}{c} A_{1,\lambda }^{in} ( E ; \chi ) \\ A_{2, \lambda }^{in} ( E ; \chi ) \\ A_{3, \lambda }^{in}( E ; \chi )  \\ 
 A_{4 , \lambda }^{in}( E ; \chi )  \end{array} \right] = \left[\begin{array}{c} A_{1 , \lambda}^-( E ; \chi )  \\
  A_{3 , \lambda}^-( E ; \chi )  \\   A_{2 , \lambda}^+ ( E ; \chi ) \\  A_{4 , \lambda}^+ ( E ; \chi ) \end{array} \right]
  \;\;\; , \;\;
 \left[\begin{array}{c} A_{1, \lambda }^{out}( E ; \chi )  \\ A_{2, \lambda }^{out}( E ; \chi )  \\ A_{3, \lambda }^{out}( E ; \chi )  \\ 
 A_{4, \lambda}^{out} ( E ; \chi ) \end{array} \right] = \left[\begin{array}{c} A_{1 , \lambda}^+ ( E ; \chi ) \\
  A_{3 , \lambda}^+( E ; \chi )  \\   A_{2 , \lambda}^- ( E ; \chi ) \\  A_{4 , \lambda}^-( E ; \chi )  \end{array} \right]
  \:\:\:\: , 
  \label{mc.7}
  \eneq
  \noindent
Thus, the $S$-matrix elements satisfy
   
\beq
\sqrt{v_{j , \lambda} } A_{j,\lambda}^{out}( E ; \chi )  = \sum_{j' =1}^4 \sum_{\lambda' = 1}^N 
[S ( E ; \chi ) ]_{(j,\lambda) , ( j' , \lambda' ) } 
\sqrt{v_{j' , \lambda'}} A_{j' , \lambda'}^{in}( E ; \chi ) 
\:\:\:\: , 
\label{mc.8}
\eneq
\noindent
 with the velocities  $v_{j, \lambda}  = \left| \frac{d E}{d \beta_{p,\lambda}} \right|$ for $j = 1,2$, 
and $v_{j, \lambda}  = \left| \frac{d E}{d \beta_{h,\lambda}} \right|$ for $j = 3,4$. 

\subsection{Bogoliubov-de Gennes equations for a long SNS junction} 
\label{models_long}

We now consider the BDG equations within a long SNS junction, such as the one 
sketched in FIg. \ref{fig_multi0} {\bf b)}. 
We assume that the central region C runs from $x = 0$ to $x = \ell$ and, consistently, that 
the lead L extends from $x = - \infty $  to $x = 0$, while the
lead R extends from $x = \ell$ to $x = \infty$.  
Letting $K$ be the number of 
open electronic channels within C, one finds that the second-quantized 
Hamiltonian for the system is given by $H = H_L + H_R + H_C + H_T$, with 
$H_L , H_R $ given in Eqs.(\ref{newvel.01}), with the integrals respectively 
computed from $ - \infty$ to 0 and from $\ell$ to $+\infty$  \cite{likar}, and  $H_C$
given by

\beq
 H_C-\mu {\cal N}_C = \int_0^\ell \: d x \: \sum_{ \rho  = 1}^K \:  \left\{ \sum_\sigma
 \Psi_{C , \rho , \sigma}^\dagger 
 ( x ) h_{\rho} ( x )  \Psi_{C , \rho , \sigma}  ( x ) \right\}
 \;\;\;\; , 
 \label{newvel.02}
\eneq
\noindent
with $ \Psi_{C , \rho , \sigma}  ( x )$ being  the annihilation operator 
for an electron in channel-$\rho$ with spin $\sigma$ within C, ${\cal N}_C$ being 
the total particle number within C, and 
$h_\rho ( x ) = - \frac{1}{2 m_\rho} \frac{d^2 }{d x^2} +V_\rho $ 
is the corresponding single-fermion  Hamiltonian, with $m_\rho$ 
being the effective electron mass and $V_\rho$ being the  potential
within channel $\rho$. In addition, in order for $H_L , H_R $  in Eqs.(\ref{newvel.01})
and $H_C$ in Eq. (\ref{newvel.02}) to be well-defined, we impose boundary
conditions on  $\Psi_{L , \lambda , \sigma}  ( x )$ and  $\Psi_{C , \rho , \sigma}  ( x )$
at $x=0$, as well as on  $\Psi_{R , \lambda , \sigma}  ( x )$ and  $\Psi_{C , \rho , \sigma}  ( x )$
at $x=\ell$, by requiring that all the derivatives with respect to $x$ vanish, so
that the fields themselves are non-zero at the interfaces.
The tunneling Hamiltonian $H_T$ encodes the coupling 
between C and the leads. We assume it to take the generic form 

\beq
H_T = \sum_{ \lambda = 1}^N \sum_{\rho = 1}^K \sum_{ \sigma = \uparrow , \downarrow} 
\{ [ t_L ]_{\lambda , \rho} \Psi_{L , \lambda,  \sigma}^\dagger ( 0 ) \Psi_{C, \rho , \sigma}  ( 0 ) 
+ {\rm h.c.} \} 
+ \sum_{ \lambda = 1}^N \sum_{\rho = 1}^K  \sum_{ \sigma = \uparrow , \downarrow} 
\{ [ t_R ]_{\lambda , \rho} \Psi_{R , \lambda,  \sigma}^\dagger ( \ell ) 
\Psi_{C , \rho , \sigma} ( \ell ) 
+ {\rm h.c.} \} 
\:\:\:\: , 
\label{sss.b1}
\eneq
\noindent
where $t_{L; \lambda , \rho} , t_{R; \lambda , \rho} $ are
tunneling amplitude matrices independent of $\sigma$.
Writing an energy eigemode operator of $H_C$, 
$\gamma_{ C, \rho , \sigma} ( E ) $, as 

\beq
\gamma_{  C , \rho  , \sigma} ( E ) = \int_{0}^\ell \: d x \: \{ u_{C , \rho ,E} ( x ) 
\Psi_{ C , \rho , \sigma} ( x ) + \sigma v_{C , \rho , E} ( x ) 
\Psi_{ C , \rho , \sigma}^\dagger  ( x ) \}
\;\;\;\; , 
\label{novelk.r}
\eneq
\noindent
  and requiring that $ [ \gamma_{  C , \rho , \sigma} ( E ) , H_C ] = 
  E  \gamma_{  C , \rho , \sigma} 
  ( E ) $ yields the BDG equations for the wavefunctions within C, $u_{C , \rho, E}  ( x ) 
, v_{C , \rho, E} ( x )$, given by
 
\begin{eqnarray}
\left[ - \frac{1}{2 m_\rho} \frac{d^2 }{d x^2} +V_\rho \right] 
 u_{C , \rho, E} ( x )  &=& 
 E u_{C , \rho , E}  ( x ) \nonumber \\
\left[   \frac{1}{2 m_\rho} \frac{d^2 }{d x^2} -V_\rho \right] 
 v_{C , \rho , E}  ( x ) &=& 
 E v_{C , \rho, E} ( x ) 
 \:\:\:\: , 
 \label{b1.1}
\end{eqnarray}
\noindent
with $0 < x < \ell$. 
A generic solution  to the Eqs.(\ref{b1.1}) can then 
be written as 

\beq
\left[ \begin{array}{c}
        u_{C , \rho , E} ( x ) \\ v_{C , \rho , E} ( x ) 
       \end{array} \right] = \sum_{ j = 1}^4 C_{j , \rho} (E ; \chi )
       \left[ \begin{array}{c}
        u_{C , \rho , E } ( x ) \\ v_{C , \rho , E} ( x ) 
       \end{array} \right]_j 
\:\:\:\: , 
\label{centr.1}
\eneq
\noindent
with 

\beq
\left[ \begin{array}{c}
        u_{C , \rho , E} ( x ) \\ v_{C , \rho , E} ( x ) 
       \end{array} \right]_{1(2)} =  \left[ \begin{array}{c}
       e^{ \pm i \alpha_{p , \rho } x }  \\ 0
       \end{array} \right]  \;\;\; , \;\;
 \left[ \begin{array}{c}
        u_{C , \rho , E} ( x ) \\ v_{C , \rho , E} ( x ) 
       \end{array} \right]_{3(4)} =  \left[ \begin{array}{c} 0 \\
       e^{ \mp i \alpha_{h , \rho } x }  
       \end{array} \right]        
\:\:\:\: , 
\label{centr.2}
\eneq
\noindent
and 

\begin{eqnarray}
 \alpha_{ p , \rho} &=& \sqrt{2 m_\rho ( -V_\rho + E ) } \nonumber \\
  \alpha_{ h , \rho} &=& \sqrt{2 m_\rho ( -V_\rho - E ) } 
  \:\:\:\: . 
  \label{centr.3}
\end{eqnarray}
\noindent
The scattering processes at the 
interfaces are determined by the specific form of $H_T$ in Eq. (\ref{sss.b1})
and are encoded in the transmission matrix from L to C, 
$L ( E )$, and in the transmission matrix from C to R, $ R ( E ) $. 
In general, $L ( E )$ and $R(E)$ are  a $4K \times 4N$ and a   
$4N \times 4K$-rectangular matrix, respectively, defined so that 

\begin{eqnarray}
 C_{j , \rho} (E ; \chi ) &=& \sum_{ j' = 1}^4 \sum_{ \lambda = 1}^N [ L ( E ; \chi ) ]_{(j , \rho) , (j' , \lambda)}  
 A_{j' , \lambda}^- (E ; \chi )\nonumber \\
  A_{j , \rho}^+(E ; \chi ) &=& \sum_{ j' = 1}^4 \sum_{ \rho = 1}^K [ R ( E ; \chi ) ]_{(j , \lambda) , (j' , \rho)}  
 C_{j' , \rho} (E ; \chi )
 \:\:\:\: .
 \label{centr.4}
\end{eqnarray}
\noindent
Once the transmission matrices at the interfaces are defined as in Eq. (\ref{centr.4}), the 
factorizability of the total transmission matrix readily yields Eq. (\ref{mm.1}) of the 
main text. 

\section{Derivation of Eqs.(\ref{mc.20},\ref{mc.25}) for a multi-channel junction}
\label{proof}

In this appendix, we  derive Eq. (\ref{mc.20}) in the multi-channel
case, together with the relation between ${\cal F} [ E ; \chi ]$, ${\cal G} [ E ; \chi ]$
and the $M$-matrix elements (Eq. (\ref{mc.25}))  (for notational simplicity, we 
will drop throughout all the appendix the dependence of the amplitudes and of 
the matrix elements on $E$ and $\chi$.) To do so, we consider a solution of 
the BDG equations for a multi-channel system discussed in Appendix \ref{models}
with boundary conditions corresponding to putting  the system in a large box, 
ranging from $x = - L/2$ to $x = L/2 + \ell$, that is, we require that the wavefunctions
are equal to 0 both  at $x = - L/2$ and at $x = L/2 + \ell$. This constrains 
the form of the solutions, leading to consistency relations between the momenta, which
can be either expressed in terms of the $M$, or of the $S$-matrix elements. Equating
corresponding quantities expressed in formally different ways, we 
eventually derive Eqs.(\ref{mc.20},\ref{mc.25}). 

Imposing vanishing boundary conditions as described above implies,
at the left-hand boundary of the box ($x  = - L/2$)
$u_{L , \lambda , E} ( x = - L/2) = 
v_{L , \lambda , E} ( x = - L/2) = 0$. Thus, 
from Eqs.(\ref{mc.4}) we obtain  

\begin{eqnarray}
  \cos \left( \frac{\Psi}{2} \right) \{ e^{  - \frac{i}{2} \beta_{p , \lambda} L} A_{1, \lambda }^- + 
  e^{ \frac{i}{2}  \beta_{p , \lambda} L} A_{2, \lambda }^- \} 
  - \sin \left( \frac{\Psi}{2} \right) e^{ \frac{i}{2} \chi }  
  \{ e^{   \frac{i}{2}  \beta_{ h , \lambda}  L} A_{3, \lambda }^- + e^{ - \frac{i}{2}  \beta_{h , \lambda} L} 
  A_{4, \lambda }^- \} &=& 0 \nonumber \\
    - \sin \left( \frac{\Psi}{2} \right) e^{ - \frac{i}{2} \chi } 
    \{ e^{  - \frac{i}{2}  \beta_{p , \lambda} L} A_{1, \lambda }^- + e^{ \frac{i}{2}  \beta_{p , \lambda} L} 
    A_{2, \lambda }^- \} + 
 \cos \left( \frac{\Psi}{2} \right)       
 \{ e^{   \frac{i}{2}  \beta_{h , \lambda}  L} A_{3, \lambda }^- + e^{ - \frac{i}{2}  \beta_{h , \lambda}  L} 
 A_{4, \lambda }^- \} &=& 0 
 \:\:\:\: ,
 \label{mc.8a}
\end{eqnarray}
\noindent
with $\lambda = 1,2, \ldots, N$. Similarly, at the 
right-hand boundary of the box, we impose
$u_{R , \lambda , E} ( x =  L/2 + \ell) = 
v_{L , \lambda , E} ( x = L/2 + \ell) = 0$. As a result, 
from Eqs.(\ref{mc.5}) we obtain 

\begin{eqnarray}
 \cos \left( \frac{\Psi}{2} \right) \{  e^{ \frac{i}{2}  \beta_{p , \lambda} L} \sum_{\lambda' = 1}^N
 \sum_{j' = 1}^4 M_{(1, \lambda ) , ( j' ,  \lambda'  ) } A_{ j' ,  \lambda' }^- + 
 e^{ - \frac{i}{2}  \beta_{p , \lambda } L} \sum_{\lambda' = 1}^N  
 \sum_{j' = 1}^4 M_{(2 , \lambda ) , ( j' ,  \lambda' ) } A_{j' , \lambda' }^- \} && \nonumber \\
 - e^{ - \frac{i}{2}  \chi } \sin  \left( \frac{\Psi}{2} \right) \{ e^{ -  \frac{i}{2}  \beta_{ h , 
 \lambda} L} \sum_{\lambda' = 1}^N   \sum_{j' = 1}^4 M_{(3, \lambda ) , (j' ,  \lambda'  ) }
 A_{j' , \lambda' }^- + e^{    \frac{i}{2}  \beta_{ h , 
 \lambda } L} 
  \sum_{\lambda' = 1}^N 
 \sum_{j' = 1}^4 M_{(4, \lambda ) , ( j ' ,  \lambda' ) } A_{j' , \lambda' }^- \} &=& 0 \nonumber \\
 - e^{  \frac{i}{2}  \chi } \sin  \left( \frac{\Psi}{2} \right) \{ 
 e^{ \frac{i}{2}  \beta_{p , \lambda} L}  \sum_{\lambda' = 1}^N 
 \sum_{j' = 1}^4 M_{(1, \lambda ) , ( j' ,  \lambda'  ) } A_{ j ' ,  \lambda' }^- + 
 e^{ - \frac{i}{2}  \beta_{p , \lambda} L}  \sum_{\lambda' = 1}^N 
 \sum_{j' = 1}^4 M_{(2, \lambda ) , ( j' , \lambda'  ) } A_{j' , \lambda' }^- \} && \nonumber \\
 +  \cos \left( \frac{\Psi}{2} \right) \{ e^{ -  \frac{i}{2}  \beta_{ h , 
 \lambda} L}  \sum_{\lambda' = 1}^N 
 \sum_{j' = 1}^4 M_{(3 , \lambda ) , ( j' , \lambda' ) } A_{ j' , \lambda' }^- + 
 e^{  \frac{i}{2}  \beta_{ h , 
 \lambda} L} \sum_{\lambda' = 1}^N 
 \sum_{j' = 1}^4 M_{(4 , \lambda ) , ( j' , \lambda'  ) } A_{ j' , \lambda' }^- \} &=& 0 
 \:\:\:\: ,
 \label{mc.9}
\end{eqnarray}
\noindent
with $\lambda = 1,2, \ldots, N$. Eqs.(\ref{mc.8a},\ref{mc.9}) can be 
regarded as a homogenous system in the $4N$ unknowns  $A_{ \lambda , j}^-$, 
which can be rewritten as 

\beq
\sum_{ j' = 1}^4 \sum_{ \lambda' = 1}^N {\cal A}_{(j,\lambda) , ( j' , \lambda' ) } 
A_{ j' , \lambda'}^- = 0 
\:\:\:\: , 
\label{mcc.a1}
\eneq
\noindent
with the matrix elements ${\cal A}_{(j,\lambda) , ( j' , \lambda' ) } $
given by

\beq
 {\cal A}_{(j,\lambda) , ( j' , \lambda' ) } = 
 \sum_{ j'' = 1}^4 \sum_{ \lambda'' = 1}^N 
 [ \delta_{ \lambda , \lambda''} {\cal M}_{ j , j''} ]
 \alpha_{ ( j'' , \lambda'' ) , ( j' , \lambda' ) } 
 \;\;\;\; , 
 \label{mcc.a2}
 \eneq
 \noindent
 with ${\cal M}$ being a 4$\times$4 matrix defined as 
 
\beq
{\cal M} = \left[ \begin{array}{cccc} 0 &  \cos \left( \frac{\Psi}{2} \right) 
& 0 &   - e^{        \frac{i}{2} \chi }  
\sin \left( \frac{\Psi}{2} \right)  \\ 0 & - e^{ - \frac{i}{2} \chi   }       \sin \left( \frac{\Psi}{2} \right)
& 0 & \cos \left( \frac{\Psi}{2} \right) \\ 
   \cos \left( \frac{\Psi}{2} \right) & 0 & -e^{ - \frac{i}{2} \chi }  \sin \left( \frac{\Psi}{2} \right) & 0  \\
-e^{  \frac{i}{2} \chi }  \sin \left( \frac{\Psi}{2} \right) & 0 
                  &    \cos \left( \frac{\Psi}{2} \right) & 0   \end{array} \right]
\:\:\:\: , 
\label{mc.ab13}
\eneq
\noindent
and 

\begin{eqnarray}
 \alpha_{ (j'' , \lambda'' ) , ( j' , \lambda' ) } &=& 
 \delta_{ j'', 1} \{ e^{ \frac{i}{2} \beta_{ p , \lambda''}L } 
 M_{(1 , \lambda'' ) , ( j' , \lambda' ) } + 
  e^{ - \frac{i}{2} \beta_{ p , \lambda''}L } 
 M_{(2 , \lambda'' ) , ( j' , \lambda' ) } \} \nonumber \\
 &+& \delta_{ j'', 2} \{ e^{ -  \frac{i}{2} \beta_{ p , \lambda''}L } 
\delta_{ j' , 1} + e^{    \frac{i}{2} \beta_{ p , \lambda''}L } 
\delta_{ j'   , 2} \} \delta_{ \lambda'' , \lambda'} \nonumber \\
&+&  \delta_{ j'' , 3 }  \{ e^{ - \frac{i}{2} \beta_{ h , \lambda''L} } 
 M_{(3 , \lambda'' ) , ( j' , \lambda' ) } + 
  e^{  \frac{i}{2} \beta_{ h , \lambda''}L } 
 M_{(4 , \lambda'' ) , ( j' , \lambda' ) } \} \nonumber \\
  &+& \delta_{ j'', 4} \{ e^{   \frac{i}{2} \beta_{ h , \lambda''}L } 
\delta_{ j'  ,3} + e^{  -  \frac{i}{2} \beta_{ h , \lambda''}L } 
\delta_{ j'   , 4} \} \delta_{ \lambda'' , \lambda'} 
\:\:\:\: . 
\label{mcc.aa.3}
\end{eqnarray}
\noindent
The consistency condition for having nonzero solutions for the amplitudes 
$A_{j , \lambda}^-$ then reads ${\rm det} \parallel  {\cal A}_{(j,\lambda) , ( j' , \lambda' ) }
\parallel = 0$, that is, $\cos^{2N} (  \Psi ) { \rm det} \parallel  
\alpha_{(j,\lambda) , ( j' , \lambda' ) } \parallel = 0 
\Rightarrow { \rm det} \parallel  
\alpha_{(j,\lambda) , ( j' , \lambda' ) } \parallel = 0 $. 
By pertinently grouping powers of $e^{ \pm i  \beta_{ p , \lambda} L } $ and 
of $e^{\mp i \beta_{h , \lambda} L}$, this latter condition 
gives rise to the equation 

\beq
c_A \prod_{ \lambda = 1}^N e^{ i [ \beta_{ p , \lambda} - \beta_{ h , \lambda} ] L } 
+ \ldots + 
c_B \prod_{ \lambda = 1}^N e^{ - i [ \beta_{ p , \lambda} - \beta_{ h , \lambda} ] L } 
= 0 
\;\;\;\; , 
\label{mcc.aa.4}
\eneq
\noindent
where we have introduced the ellipses to represent terms
$\propto  \prod_{ \lambda = 1}^N 
e^{ i [ a_\lambda \beta_{ p , \lambda} - b_\lambda \beta_{ h , \lambda} ] L }$, 
with $a_\lambda , b_\lambda = \pm1$ and 
at least one of the $a_\lambda$ and/or $b_\lambda$ different from the others.
It is, now, clear that $c_A \prod_{ \lambda = 1}^N e^{ i [ \beta_{ p , \lambda} - \beta_{ h , \lambda} ] L } $
is given by the determinant of the matrix obtained from $\parallel  
\alpha_{(j,\lambda) , ( j' , \lambda' ) } \parallel $ by setting to 0
all the contributions not proportional to either $e^{ \frac{i}{2} \beta_{p , \lambda} L}$, 
or to $e^{ - \frac{i}{2} \beta_{h , \lambda} L}$, that is, one obtains

\beq
c_A \prod_{ \lambda = 1}^N e^{ i [ \beta_{ p , \lambda} - \beta_{ h , \lambda} ] L } 
= {\rm det} \parallel \alpha^A_{(j, \lambda )  , (j' , \lambda')} \parallel
\;\;\;\; , 
\label{mcc.aa.5}
\eneq
\noindent
with

\begin{eqnarray}
 \alpha^A_{ (j , \lambda ) , ( j' , \lambda' ) } &=& 
  e^{ \frac{i}{2} \beta_{ p , \lambda}L } \{\delta_{ j, 1}  
 M_{(1 , \lambda ) , ( j' , \lambda' ) }  +\delta_{ j, 2}  \delta_{ j' , 2} \delta_{ \lambda  , \lambda'} \}  \nonumber \\
&+&   e^{ - \frac{i}{2} \beta_{ h , \lambda } L } \{\delta_{ j , 3 }  
 M_{(3 , \lambda ) , ( j' , \lambda' ) } 
+  \delta_{ j , 4}   
\delta_{j'  , 4} \delta_{ \lambda  , \lambda'} \} 
\:\:\:\: . 
\label{mcc.aa.6}
\end{eqnarray}
\noindent
This can be rewritten as the matrix product of a diagonal matrix containing all the $\beta$-dependence 
and the matrix $M^A$ defined in  Eq. (\ref{mc.18}):

\beq
 \alpha^A_{ (j , \lambda ) , ( j' , \lambda' ) } = \sum_{ j''=1}^4 \sum_{ \lambda'' = 1}^N 
 \{ e^{ \frac{i}{2} \beta_{ p , \lambda} L} \delta_{ \lambda , \lambda''} \delta_{j , j''} [ 
 \delta_{ j , 1} + \delta_{ j , 2 } ] + 
  e^{ - \frac{i}{2} \beta_{ h , \lambda} L} \delta_{ \lambda , \lambda''} \delta_{j , j''} [ 
 \delta_{ j , 3} + \delta_{ j , 4 } ]  \} [ M^A ]_{ ( j'' , \lambda'') , (j' , \lambda' ) }
 \:\:\:\: .
 \label{kk.a1}
\eneq
\noindent
This shows that
\beq
c_A = {\rm det} [M^A ] 
\:\:\:\: . 
\label{kk.a2}
\eneq
\noindent
Going through similar arguments, one readily proves that 
\beq
c_B \prod_{ \lambda = 1}^N e^{   i [ \beta_{ p , \lambda} - \beta_{ h , \lambda} ] L } 
= {\rm det} \parallel \alpha^B_{(j, \lambda )  , (j' , \lambda')} \parallel
\;\;\;\; , 
\label{mcc.aa.7}
\eneq
\noindent
with 
\begin{eqnarray}
 \alpha^B_{ (j , \lambda  ) , ( j' , \lambda' ) } &=& 
  e^{ - \frac{i}{2} \beta_{ p , \lambda }L } \{\delta_{ j ,  1} 
 M_{(2 , \lambda  ) , ( j' , \lambda' ) }+  \delta_{ j  , 2}   
\delta_{ j' , 1}  \delta_{ \lambda  , \lambda'}  \}  \nonumber \\
&+&  
  e^{  \frac{i}{2} \beta_{ h , \lambda }L } \{ \delta_{ j  , 3 }  
 M_{(4 , \lambda  ) , ( j' , \lambda' ) }+  \delta_{ j , 4}   
\delta_{ j'  ,3} \delta_{ \lambda  \lambda'}  \}
\:\:\:\: . 
\label{mcc.aa.8}
\end{eqnarray}
\noindent
A factorization similar to the one in Eq. (\ref{kk.a1}) takes place in 
this case, as well, in the form 

\beq
 \alpha^B_{ (j , \lambda ) , ( j' , \lambda' ) } = \sum_{ j''=1}^4 \sum_{ \lambda'' = 1}^N 
 \{ e^{ - \frac{i}{2} \beta_{ p , \lambda} L} \delta_{ \lambda , \lambda''} \delta_{j , j''} [ 
 \delta_{ j , 1} + \delta_{ j , 2 } ] + 
  e^{  \frac{i}{2} \beta_{ h , \lambda} L} \delta_{ \lambda , \lambda''} \delta_{j , j''} [ 
 \delta_{ j , 3} + \delta_{ j , 4 } ]  \} [ M^B ]_{ ( j'' , \lambda'') , (j' , \lambda' ) }
 \:\:\:\: ,
 \label{kk.a3}
\eneq
\noindent
with $M^B$ given in Eq. (\ref{mc.19}), 
which implies
\beq
c_B = {\rm det} [M^B ] 
\:\:\:\: . 
\label{kk.a4}
\eneq
\noindent
 As a result,  we then see that Eq. (\ref{mcc.aa.4}) can be recast in the form 

\beq
\left[ \frac{{\rm det} [ M^A]}{{\rm det} [ M^B] } \right] \: \prod_{ \lambda = 1}^N \: e^{ 2 i [ \beta_{ p , \lambda} -
\beta_{h , \lambda} ] L } \: + \ldots + 1 = 0 
\:\:\:\: . 
\label{mcc.aa.9}
\eneq
\noindent
To relate  $\frac{{\rm det} [ M^A]}{{\rm det} [ M^B] } $ to the determinant of the 
$S$-matrix, we use Eq. (\ref{mc.8}) to trade Eqs.(\ref{mc.8a},\ref{mc.9}) for an
algebraic system of $4N$-equations in the unknowns $A_{j , \lambda}^{  in}$. The 
resulting system is

\begin{eqnarray}
&& \cos \left( \frac{\Psi}{2} \right) \left\{ e^{ - \frac{i}{2} \beta_{ p , \lambda} L } 
A_{1,\lambda}^{in} + e^{  \frac{i}{2} \beta_{ p , \lambda} L } 
\sum_{ j' = 1}^4 \sum_{\lambda' = 1}^N \sqrt{\frac{v_{j', \lambda'}}{v_{3 , \lambda}}}
S_{ (3, \lambda) , (j' , \lambda' ) } A_{j' , \lambda'}^{in} \right\} \nonumber \\
&-& e^{ \frac{i}{2} \chi}  \sin \left( \frac{\Psi}{2} \right) \left\{  
e^{  \frac{i}{2} \beta_{ h , \lambda} L } A_{2, \lambda}^{in} + 
e^{ -   \frac{i}{2} \beta_{ h , \lambda} L }\sum_{ j' = 1}^4 \sum_{\lambda' = 1}^N 
\sqrt{\frac{v_{j', \lambda'}}{v_{4 , \lambda}}}
S_{ (4, \lambda) , (j' , \lambda' ) } A_{j', \lambda'}^{in} \right\} = 0 \nonumber \\
&-& e^{ -\frac{i}{2} \chi}  \sin \left( \frac{\Psi}{2} \right) \left\{  
e^{ - \frac{i}{2} \beta_{p , \lambda} L } A_{1, \lambda}^{in} + 
e^{    \frac{i}{2} \beta_{ p , \lambda} L }\sum_{ j' = 1}^4 \sum_{\lambda' = 1}^N 
\sqrt{\frac{v_{j', \lambda'}}{v_{3 , \lambda}}}
S_{ (3, \lambda) , (j' , \lambda' ) } A_{j', \lambda'}^{in} \right\} \nonumber \\
&+&  \cos \left( \frac{\Psi}{2} \right) \left\{ e^{  \frac{i}{2} \beta_{h , \lambda} L } 
A_{2,\lambda}^{in} + e^{-  \frac{i}{2} \beta_{ h , \lambda} L } 
\sum_{ j' = 1}^4 \sum_{\lambda' = 1}^N \sqrt{\frac{v_{j', \lambda'}}{v_{4 , \lambda}}}
S_{ (4, \lambda) , (j' , \lambda' ) } A_{j' , \lambda'}^{in} \right\} = 0 \nonumber \\
&& \cos \left( \frac{\Psi}{2} \right) \left\{ e^{ \frac{i}{2} \beta_{ p , \lambda} L } 
\sum_{ j' = 1}^4 \sum_{\lambda' = 1}^N \sqrt{\frac{v_{j', \lambda'}}{v_{1 , \lambda}}}
S_{ (1, \lambda) , (j' , \lambda' ) } A_{j' , \lambda'}^{in} +
 e^{ - \frac{i}{2} \beta_{ p , \lambda} L } A_{3,\lambda}^{in}  
\right\} \nonumber \\
&-&e^{ - \frac{i}{2} \chi}  \sin \left( \frac{\Psi}{2} \right) \left\{  
e^{  - \frac{i}{2} \beta_{ h , \lambda} L }\sum_{ j' = 1}^4 \sum_{\lambda' = 1}^N 
\sqrt{\frac{v_{j', \lambda'}}{v_{2 , \lambda}}}
S_{ (2, \lambda) , (j' , \lambda' ) } A_{j', \lambda'}^{in} 
+ e^{  \frac{i}{2} \beta_{ h , \lambda} L }
A_{4, \lambda}^{in}  
\right\} = 0 \nonumber \\
&-& e^{  \frac{i}{2} \chi}  \sin \left( \frac{\Psi}{2} \right) \left\{  
e^{  \frac{i}{2} \beta_{p , \lambda} L }\sum_{ j' = 1}^4 \sum_{\lambda' = 1}^N 
\sqrt{\frac{v_{j', \lambda'}}{v_{1 , \lambda}}}
S_{ (1, \lambda) , (j' , \lambda' ) } A_{j', \lambda'}^{in}
+e^{ -   \frac{i}{2} \beta_{ p , \lambda} L }
A_{3, \lambda}^{in}  
 \right\} \nonumber \\
&+&  \cos \left( \frac{\Psi}{2} \right) \left\{ e^{ - \frac{i}{2} \beta_{h , \lambda} L } 
\sum_{ j' = 1}^4 \sum_{\lambda' = 1}^N \sqrt{\frac{v_{j', \lambda'}}{v_{2 , \lambda}}}
S_{ (2, \lambda) , (j' , \lambda' ) } A_{j' , \lambda'}^{in} +e^{  \frac{i}{2} \beta_{ h , \lambda} L } 
A_{4,\lambda}^{in}  \right\} = 0 
\:\:\:\: . 
\label{mcc.aa.10}
\end{eqnarray}
\noindent
As $\lambda = 1, \ldots , N$, Eqs.(\ref{mcc.aa.10}) define a 
$4N$-equation system in the unknowns $A_{j , \lambda }^{in}$,
which can be rewritten as 
 
\beq
\sum_{ j' = 1}^4 \sum_{ \lambda' = 1}^N {\cal B}_{(j,\lambda) , ( j' , \lambda' ) } 
A_{ j' , \lambda'}^{in} = 0 
\:\:\:\: , 
\label{mcc.aa.11}
\eneq
\noindent
with the matrix elements ${\cal B}_{(j,\lambda) , ( j' , \lambda' ) } $
given by

\beq
 {\cal B}_{(j,\lambda) , ( j' , \lambda' ) } = 
 \sum_{ j'' = 1}^4 \sum_{ \lambda'' = 1}^N 
 [ \delta_{ \lambda , \lambda''} {\cal M}_{ j , j''} ]
 \beta_{ ( j'' , \lambda'' ) , ( j' , \lambda' ) } 
 \;\;\;\; , 
 \label{mcc.aa.12}
 \eneq
 \noindent
 and
 
 \begin{eqnarray}
  \beta_{ ( j , \lambda ) , (j' , \lambda') } &=& \delta_{j , 1} \left\{ 
  e^{ \frac{i}{2} \beta_{p , \lambda} L} \sqrt{\frac{v_{j' , \lambda'} }{v_{1 , \lambda}}}
  S_{(1 , \lambda) , ( j' , \lambda' ) } +  e^{ - \frac{i}{2} \beta_{p , \lambda} L}
  \delta_{j' , 3 } \delta_{ \lambda , \lambda'} \right\} \nonumber \\
  &+& \delta_{j,2} \left\{  e^{ - \frac{i}{2} \beta_{p , \lambda} L} \delta_{j' ,1 } 
  \delta_{ \lambda , \lambda'} +  e^{ \frac{i}{2} \beta_{p , \lambda} L}
  \sqrt{\frac{v_{j' , \lambda'} }{v_{3 , \lambda}}}
  S_{(3 , \lambda) , ( j' , \lambda' ) }  \right\} \nonumber \\
  &+& \delta_{j,3} \left\{   e^{ - \frac{i}{2} \beta_{h , \lambda} L}   
  \sqrt{\frac{v_{j' , \lambda'} }{v_{2 , \lambda}}}
  S_{(2 , \lambda) , ( j' , \lambda' ) } +  e^{   \frac{i}{2} \beta_{h , \lambda} L}   
    \delta_{j' , 4} \delta_{ \lambda , \lambda'}  \right\} \nonumber \\
    &+&  \delta_{j,4} \left\{ e^{   \frac{i}{2} \beta_{h , \lambda} L} \delta_{j' ,2 } 
  \delta_{ \lambda , \lambda'} +  e^{- \frac{i}{2} \beta_{h , \lambda} L}
  \sqrt{\frac{v_{j' , \lambda'} }{v_{4 , \lambda}}}
  S_{(4 , \lambda) , ( j' , \lambda' ) }  \right\} 
  \:\:\:\: . 
  \label{mcc.aa.13}
 \end{eqnarray}
\noindent
The consistency condition for having nonzero solutions therefore reads 
${\rm det} \parallel {\cal B}_{( j , \lambda ) , (j' , \lambda' ) } \parallel = 
\cos^{2N} ( \Psi ) {\rm det} \parallel \beta_{( j , \lambda ) , (j' , \lambda' ) } \parallel 
= 0 $, which implies  ${\rm det} \parallel \beta_{( j , \lambda ) , (j' , \lambda' ) } \parallel 
= 0 $. As we have done before, by pertinently grouping powers of $e^{ \pm  i \beta_{ p , \lambda} L }$
and of $e^{ \mp i \beta_{ h , \lambda} L }$, we trade the condition on the determinant 
for the equivalent equation 

\beq
\delta_A \prod_{ \lambda = 1}^N e^{ i [ \beta_{p , \lambda} - \beta_{ h , \lambda } ] L } + \ldots 
+ \delta_B \prod_{ \lambda = 1}^N e^{ - i [ \beta_{p , \lambda} - \beta_{ h , \lambda } ] L }  = 0 
\:\:\:\: . 
\label{mcc.aa.14}
\eneq
\noindent
As we have done before, we therefore
compute $\delta_A \prod_{ \lambda = 1}^N e^{ i [ \beta_{p , \lambda} - \beta_{ h , \lambda } ] L }$
as $\delta_A \prod_{ \lambda = 1}^N e^{ i [ \beta_{p , \lambda} - \beta_{ h , \lambda } ] L } = 
{\rm det} \parallel \beta^A_{(j , \lambda ) , (j' , \lambda' ) }\parallel$, with 

\begin{eqnarray}
  \beta^A_{(j , \lambda ) , (j' , \lambda' ) } &=& e^{ \frac{i}{2} \beta_{p , \lambda} L } \left\{ 
  \delta_{j,1} \sqrt{\frac{v_{j' , \lambda'} }{v_{1 , \lambda}}}
  S_{(1 , \lambda) , ( j' , \lambda' ) }  + \delta_{j,2} \sqrt{\frac{v_{j' , \lambda'} }{v_{3 , \lambda}}}
  S_{(3 , \lambda) , ( j' , \lambda' ) } \right\} \nonumber \\
 &+& e^{ - \frac{i}{2} \beta_{h , \lambda} L } \left\{ \delta_{j,3}  \sqrt{\frac{v_{j' , \lambda'} }{v_{2 , \lambda}}}
  S_{(2 , \lambda) , ( j' , \lambda' ) } + 
\delta_{j,4}
\sqrt{\frac{v_{j' , \lambda'} }{v_{4 , \lambda}}}  S_{(4 , \lambda) , ( j' , \lambda' ) } \right\} 
  \;\;\;\; . 
  \label{mcc.aa.15}
  \end{eqnarray}
\noindent
At variance, we obtain  $\delta_B \prod_{ \lambda = 1}^N e^{ - i [ \beta_{p , \lambda} - \beta_{ h , \lambda } ] L } = 
{\rm det} \parallel \beta^B_{(j , \lambda ) , (j' , \lambda' ) }\parallel$, with 

\begin{eqnarray}
  \beta^B_{(j , \lambda ) , (j' , \lambda' ) } &=&  e^{ - \frac{i}{2} \beta_{p , \lambda} L } \left\{ 
    \delta_{j,1}  \delta_{j' , 3 } \delta_{ \lambda , \lambda'}  +  \delta_{j,2}  \delta_{j' ,1 } 
  \delta_{ \lambda , \lambda'} \right\} \nonumber \\
  &+&  e^{   \frac{i}{2} \beta_{h , \lambda} L } \left\{ \delta_{j,3}
  \delta_{j' , 4} \delta_{ \lambda , \lambda'} + \delta_{j,4}
  \delta_{j' , 2} \delta_{ \lambda , \lambda'} \right\}
\:\:\:\: . 
\label{mcc.aa.16}
\end{eqnarray}
\noindent
Thus, we obtain 

\begin{eqnarray}
 \delta_A   &=& (-1)^N \: {\rm det} [ S ] \nonumber \\
 \delta_B &=& (-1)^N
 \:\:\:\: . 
 \label{mcc.aa.17}
\end{eqnarray}
\noindent
 As a  consequence of Eqs.(\ref{mcc.aa.17}), we see that Eq. (\ref{mcc.aa.14}) 
 can be recast in the form 
 
\beq
{\rm det} [ S ] \: \prod_{\lambda = 1}^N e^{ 2 i [ \beta_{p , \lambda} - \beta_{h , \lambda} ] L } 
+ \ldots + 1 = 0 
\;\;\;\; . 
\label{mcc.aa.18}
\eneq
\noindent
Though Eqs.(\ref{mcc.aa.9},\ref{mcc.aa.18}) have been obtained following
two alternative routes, they must clearly coincide with each other, once
the coefficients are consistently normalized, as we did. As a result, the coefficients
of $\prod_{\lambda = 1}^N e^{ 2 i [ \beta_{p , \lambda} - \beta_{h , \lambda} ] L } $
must be equal to each other, which implies  
Eqs.(\ref{mc.20},\ref{mc.25}) of the main text.

\section{Derivation of Eq. (\ref{sol.12})}
\label{root_poly}

Eq. (\ref{sol.12}) is one of the key results of this paper, as it
provides us with a closed-form formula to exactly expressing $I  [ \chi ; T=0 ]$
in the case of equivalent channels.  To derive Eq. (\ref{sol.12}), we
start from the result in Eq. (\ref{ss.2})
and from the observation that, based on general properties of the transmission matrix elements, as well as 
on the explicit calculation of ${\cal P} ( u ; \chi  )$, one obtains that  
$P_0 ( \chi ) = P_{2K} ( \chi ) = 1$. As a first intermediate step, let us 
define $z = - \frac{2 \omega \ell}{v}$, so that Eq. (\ref{ss.2}) becomes

\beq
I  [ \chi ; T=0 ] = \frac{2 e v}{4 \pi \ell} \: \int_{- \infty}^\infty \: d z \: 
\left[ 
\frac{\sum_{ j = - K+1}^{K-1} \partial_\chi P_{j+K} ( \chi ) 
( e^z )^j }{\sum_{ j = - K}^{K}  P_{j+K} ( \chi ) 
( e^z )^j} \right]  
\;\;\;\; . 
\label{ss.a3}
\eneq
\noindent
Next, let us multiply the numerator and the denominator of Eq. (\ref{ss.a3})
by $e^{ K z}$. We then obtain

\beq
I  [ \chi ; T=0 ] = \frac{2 e v}{4 \pi \ell} \: \int_{- \infty}^\infty \: d z \: 
\left[ 
\frac{\sum_{ j = - K+1}^{K-1} \partial_\chi P_{j+K} ( \chi ) 
( e^z )^{j+K} }{\sum_{ j = - K}^{K}  P_{j+K} ( \chi ) 
( e^z )^{j+K}} \right]    
\;\;\;\; . 
\label{ss.a4}
\eneq
\noindent
Finally, let us define $u \equiv e^z$ and use $u$ as integration variable.
This implies 

\beq
I  [ \chi ; T=0 ] = \frac{2 e v}{4 \pi \ell} \: \int_{0}^\infty \: \frac{d u}{u} \: 
\left[ 
\frac{\sum_{ j = - K+1}^{K-1} \partial_\chi P_{j+K} ( \chi ) 
u^{j+K} }{\sum_{ j = - K}^{K}  P_{j+K} ( \chi ) 
u^{j+K}} \right]    
\;\;\;\; . 
\label{ss.a5}
\eneq
\noindent
On introducing the polynomial ${\cal P} ( u  ; \chi ) = \sum_{ j = 0}^{2K} P_j ( \chi ) 
u^j  = 1 + u^{2K} + \sum_{ j = 1}^{2K-1} P_j ( \chi ) 
u^j $, with, in general, $P_j ( \chi ) \neq 1$, for $j = 1 , \ldots , 2K-1$, 
Eq. (\ref{ss.a5}) can be rewritten as 

\beq
I  [ \chi ; T=0 ] = \frac{2 e v}{4 \pi \ell} \: \int_{0}^\infty \: \frac{d u}{u} \: 
\left[ 
\frac{\partial_\chi {\cal P} ( u )  }{ {\cal P} ( u ) } \right]     =  
\frac{2 e v}{4 \pi \ell} \: \int_{0}^\infty \: \frac{d u}{u} \: \partial_\chi 
\ln [ \prod_{ j = 1}^{2K} ( u - u_j ( \chi )) ]
\;\;\;\; , 
\label{ss.a6}
\eneq
\noindent
with $u_j ( \chi )$ being the roots of $ {\cal P} ( u ; \chi  ) = 0 $.
Eq. (\ref{ss.a6}) can then be rewritten as

\beq
I  [ \chi ; T=0 ] = \frac{2 e v}{4 \pi \ell} \: \sum_{ j = 1}^{2K} 
\: \int_{0}^\infty \: \frac{d u}{u} \: \frac{\partial_\chi u_j ( \chi ) }{
(u_j ( \chi ) - u )}
\:\:\:\: . 
\label{ss.a7}
\eneq
\noindent
The argument of the integral in Eq. (\ref{ss.a7}) looks like it diverges 
as $u^{-1}$ as $u \to 0$. However, the integral is convergent, 
due to the condition  $\prod_{ j =1}^{2K} u_j ( \chi ) = 1$,
which  implies $\sum_{ j = 1}^{2K} \ln u_j ( \chi ) = 0$. 
To evidence this, we introduce a scale  $\epsilon$ to control 
the small-$u$ divergence and (though it is not strictly necessary),
a cutoff $\Lambda$ to keep under control the behavior of the 
integral in the large-$u$ region. This means that we rewrite
Eq. (\ref{ss.a7}) as 

\beq
I  [ \chi ; T=0 ] = \frac{2 e v}{4 \pi \ell} 
\: \lim_{ \Lambda \to \infty} \lim_{\epsilon \to 0 } \: \sum_{ j = 1}^{2K} \: 
\int_{\epsilon}^\Lambda \: \frac{d u}{u} \: \frac{\partial_\chi u_j ( \chi ) }{
(u_j ( \chi ) - u )}
\:\:\:\: . 
\label{ss.aa7}
\eneq
\noindent
Computing the integrals at finite cutoffs and eventually
getting rid of the cutoffs by sending $\epsilon \to 0$ and $\Lambda \to \infty$, 
by  using the relations between
the roots listed above, one obtains 

\beq
I  [ \chi ; T=0 ] = \frac{2 e v}{4 \pi \ell} \: \sum_{ j = 1}^{2K} \{\ln u_j ( \chi ) \partial_\chi 	\ln
u_j ( \chi )  \} = \frac{  e v}{ 4 \pi \ell} \: \sum_{ j = 1}^{2K} 
\partial_\chi  \ln^2 [ u_j ( \chi ) ]
\:\:\:\: . 
\label{ss.a10}
\eneq
\noindent
In the specific case $K=1$, which was considered in Ref. [\onlinecite{giu_af}], 
we obtain (using $u_2 ( \chi ) = 1 / u_1 ( \chi ) \Rightarrow \partial_\chi \{ 
\ln    u_1 ( \chi ) ]
 + \ln [ u_2 ( \chi ) ] \} = 0$)

\begin{eqnarray}
&&  \partial_\chi \ln^2 [ u_1 ( \chi ) ] + \partial_\chi \ln^2 [ u_2 ( \chi ) ] = 
\nonumber \\
&& \frac{1}{2} \partial_\chi \{ \ln [ u_1 ( \chi ) ]
 + \ln [ u_2 ( \chi ) ]  \}^2 + 
  \frac{1}{2} \partial_\chi \{ \ln [ u_1 ( \chi ) ]
- \ln [ u_2 ( \chi ) ]  \}^2 \nonumber \\
&& =  \frac{1}{2} \ln^2 \left( \frac{u_1 ( \chi ) }{u_2 ( \chi ) } \right) 
\:\:\:\: . 
\label{ss.a11}
\end{eqnarray}
\noindent
From Eq. (\ref{ss.a11}) we eventually get, for $K=1$, 

\beq
I  [ \chi ; T=0 ] = \frac{ e v}{4 \pi \ell} \: \partial_\chi 
 \ln^2 \left( \frac{u_1 ( \chi ) }{u_2 ( \chi ) } \right) 
\:\:\:\: . 
\label{ss.a12}
\eneq
\noindent
From Eq. (\ref{ss.a12}), setting 

\begin{eqnarray}
 u_1 ( \chi ) &=& e^{ i \vartheta ( \chi ) } \nonumber \\
  u_2 ( \chi ) &=& e^{ - i \vartheta ( \chi ) }
  \;\;\;\; , 
  \label{ss.a13}
\end{eqnarray}
\noindent
which implies

\beq
\frac{u_1 ( \chi ) }{u_2 ( \chi ) } = e^{ 2 i \vartheta ( \chi ) }
\:\:\:\: , 
\label{ss.a14}
\eneq
\noindent
one obtains Eq. (3) of  Ref. [\onlinecite{giu_af}].

\section{Construction of the polynomial  ${\cal P} ( u ; \chi  ) $ }
\label{co_put}

In this appendix, we work out the algorithm to explicitly construct the polynomial  ${\cal P} ( u ; \chi) $ 
we introduce in Section \ref{long_junction} to fully characterize the formula for the dc Josephson 
current in the symmetric case. In particular, we first construct ${\cal P} ( u ; \chi)$ in
full generality, that is, for generic $N$ and $K$, by expressing it as 
a function of the reflection matrices at the interfaces evaluated at the 
Fermi level. As a specific example, we then provide the explicit 
formula for $N=1$ and $K$ generic, by showing that, for $N=1$,  any system with 
$K (\geq 2)$ equivalent channels within C can be reduced to the one 
with $K=2$. 

The starting point is that, as  $| E | < \Delta$,  there are no transmitted waves outside of C. This means that,
within the left-hand lead, there will be no $\left[ \begin{array}{c} u_{L , \lambda , E} ( x ) \\
                                                            v_{L , \lambda , E} ( x ) 
                                                           \end{array} \right]_{1 , 3}$-solutions, 
while   $\left[ \begin{array}{c} u_{L , \lambda , E} ( x ) \\
                                                            v_{L , \lambda , E} ( x ) 
                                                           \end{array} \right]_{2,4}$,                                                           
will behave as evanescent waves, as $x \to - \infty$. As a result, 
we obtain $2K$ linear relations between  the coefficients of the 
solution to   Eq. (\ref{centr.1}). To formally express them, we introduce   
  the $2K \times 2K$ reflection matrix at the left-hand interface, 
  $\parallel [ R_L ( E ; \chi ) ]_{ ( a , \rho ) , (a' , \rho' ) } \parallel $, 
 with $a , a' = 1,2$ and $\rho , \rho' = 1 , \ldots ,K$, 
 such that 

\beq
C_{2a-1, \rho} = \sum_{ a' = 1,2} \sum_{ \rho' = 1}^K \sqrt{\frac{v^C_{a' , \rho'}}{v^C_{a , \rho}}}
R_L ( E ; \chi ) ]_{ ( a , \rho ) , (a' , \rho' ) } C_{2a' , \rho'} 
\:\:\:\: , 
\label{sol.15}
\eneq
\noindent
with $v^C_{1 , \rho} = \left| \frac{d E }{d \alpha_{p , \rho}} \right|$
and $v^C_{2 , \rho} = \left| \frac{d E }{d \alpha_{h , \rho}} \right|$. 
Similarly, within the right-hand lead, there will be no 
$\left[ \begin{array}{c} u_{L , \lambda , E} ( x ) \\
                                                            v_{L , \lambda , E} ( x ) 
                                                           \end{array} \right]_{2 , 4}$-solutions, 
while   $\left[ \begin{array}{c} u_{L , \lambda , E} ( x ) \\
                                                            v_{L , \lambda , E} ( x ) 
                                                           \end{array} \right]_{1,3}$,                                                           
will behave as evanescent waves, as $x \to  \infty$. This allows fow 
deriving $2K$ additional linear relations between  the coefficients of the 
solution to   Eq. (\ref{centr.1}), in terms of the $2K \times 2K$ reflection matrix 
at the right-hand interface, 
$\parallel [ R_R ( E ; \chi ) ]_{ ( a , \rho ) , (a' , \rho' ) } \parallel $, 
 such that 
 
\beq
C_{2a , \rho} = e^{ i \alpha_{ \rho}^a \ell} \sum_{ a' = 1 ,2 }
\sum_{ \rho' = 1}^K\sqrt{\frac{v^C_{a',  \rho'}}{v^C_{a , \rho}}}
[ R_R ( E ; \chi ) ]_{ ( a , \rho ) , (a' , \rho' ) }
e^{ i \alpha^{ a'}_{ \rho' } \ell } C_{ 2 a' - 1 , \rho } 
\;\;\;\; , 
\label{ssol.1}
\eneq
\noindent
with  
$\alpha^{ 1 }_{ \rho } = \alpha_{ p , \rho }$ and 
$\alpha^{ 3 }_{ \rho } = - \alpha_{ h , \rho }$. Putting together Eqs.(\ref{sol.15},\ref{ssol.1}), 
one obtains a homogeneous equation for $C_{2, \rho} , C_{4 , \rho}$, given by

\beq
\sum_{ a' = 1 , 2 } \sum_{ \rho' = 1}^K \Biggl\{ 
\delta_{a , a'} \delta_{ \rho , \rho'} - e^{ i \alpha^{a }_{ \rho} \ell} 
\sqrt{\frac{v^C_{a' , \rho'}}{v^C_{a , \rho}}}\sum_{ a'' = 1 ,2 }\sum_{ \rho'' = 1}^K
[ R_R ( E ; \chi ) ]_{ ( a , \rho ) , (a'' , \rho'' ) }
e^{ i \alpha^{ a'' }_{\rho'' } \ell }
[ R_L ( E ; \chi ) ]_{ ( a'' , \rho'' ) , (a' , \rho' ) } \Biggr\} C_{ 2 a' , \rho' } = 0 
\:\:\:\: .
\label{ssol.2}
\eneq
\noindent
In order to obtain nontrivial solutions to the system of equations reported in 
Eq. (\ref{ssol.2}), the consistency condition

\beq
{\rm det} \parallel  \delta_{a , a'} \delta_{ \rho , \rho'} - e^{ i \alpha^{a}_{ \rho} \ell} 
\sqrt{\frac{v^C_{a' , \rho'}}{v^C_{a , \rho}}}\sum_{ a'' = 1 ,2 }\sum_{ \rho'' = 1}^K
[ R_R ( E ; \chi ) ]_{ ( a , \rho ) , (a'' , \rho'' ) }
e^{ i \alpha^{a''}_{\rho'' } \ell }
[ R_L ( E ; \chi ) ]_{ ( a'' , \rho'' ) , (a' , \rho' ) } \parallel = 0 
\;\;\;\; , 
\label{ssol.3}
\eneq
\noindent
must be imposed. Eq. (\ref{ssol.3}) is the secular equation for the 
energies of the Andreev states localized within C. Restricting ourselves
to the symmetric case, we therefore assume that $\alpha^{ a}_{ \rho }$ is 
independent of $\rho$: $\alpha^{ a }_{ 1 } = \ldots = \alpha^{ a }_{ K } 
\equiv \alpha^a$. To recover the long junction limit, we then substitute $e^{ i \alpha^1 \ell} $ with
$e^{ i \alpha_F \ell} u^{-\frac{1}{2}}$ and $e^{ i \alpha^2 \ell} $ with
$e^{ - i \alpha_F \ell} u^{- \frac{1}{2}}$. This allows us to use, from now on,
the compact notation $e^{ i \alpha^a \ell} \approx 
\delta_ { a , b } [ e^{ i \sigma^z \alpha_F \ell} ]_{a, b}  u^{ - \frac{1}{2}}$, 
with $\sigma^z$ beign the third Pauli matrix. 
In addition (which amounts to neglecting  to subleading powers of $\ell^{-1}$, 
see Ref. [\onlinecite{giu_af}] for a detailed discussion), we set  $E=0$ in
the  matrices  $R_L ( E ; \chi)$ and  $R_R ( E ; \chi)$ and in the 
quasiparticle velocities $v^C_{a,\rho}$. In particular, this latter 
approximation, together with the fact that we are assuming that the 
Fermi velocities are independent of $\rho$, implies 
$\frac{v_{a' , \rho'}^C}{v_{a,\rho}^C} = 1$, independently of $a , a'$. Once the
approximations  described above have been performed,  Eq. (\ref{ssol.3}) must coincide with 
${\cal P} ( u ; \chi) = 0$, provided the normalization of the coefficients in 
the two of them has been properly chosen. As a result, multiplying Eq. (\ref{ssol.3}) by
$u^K$,  we eventually get 

\beq
{\cal P} ( u ; \chi) =  \parallel u  \delta_{a , a'} \delta_{ \rho , \rho'} -
[ e^{ i \sigma^z  \alpha_F  \ell} ]_{a , a} \sum_{ a'' = 1 ,2 }\sum_{ \rho'' = 1}^K
[ R_R ( 0 ; \chi ) ]_{ ( a , \rho ) , (a'' , \rho'' ) }
[ e^{ i \sigma^z  \alpha_F  \ell} ]_{a'' , a''} 
[ R_L ( 0 ; \chi ) ]_{ ( a'' , \rho'' ) , (a' , \rho' ) } \parallel
\:\:\:\: . 
\label{sol.9}
\eneq
\noindent
An important remark is that  Eq. (\ref{sol.9})  implies $P_0(\chi )=P_{2K}(\chi )=1$ 
since, as a general property of the solutions of the Bogoliubov - de Gennes equations, one
has that ${\rm det} [  R_L ( 0 ; \chi) ] = {\rm det} [  R_R ( 0 ; \chi) ] = 1$.

 As 
a specific example of application of Eq. (\ref{sol.9}), we now consider 
the case $N=1$. As we are going to argue next, $N=1$ is special, in
that any system with $K \geq 2$ can be traced out to a
unitary equivalent one with $K=2$. To work out the 
formula for the current in this case,   let us 
consider the tunneling Hamiltonian in Eq. (\ref{sss.b1}) in the specific case $N=1$.
 Defining $t_{L (R)} = \sqrt{\sum_{ \rho = 1}^K ( [t_{ L (R)} ]_\rho )^2}$ and 
$\vec{t}_{L(R)} = ( [t_{L(R)}]_1 , \ldots , [ t_{L (R)} ]_K )_t$, we now rotate, at fixed spin 
polarization $\sigma$, the fields 
$ \Psi_{ C , \rho , \sigma} ( x )$ by means of an unitary transformation 
$ U \in U(K)$: 
\beq
\left[ \begin{array}{c}
\tilde{\Psi}_{C , 1, \sigma} (x ) \\ \tilde{\Psi}_{C , 2, \sigma} (x ) \\
\vdots \\  \tilde{\Psi}_{C , K, \sigma} (x ) 
       \end{array} \right] = U^\dagger \cdot \left[ \begin{array}{c}
       \Psi_{C , 1, \sigma} (x ) \\ \Psi_{C , 2, \sigma} (x ) \\
\vdots \\  \Psi_{C , K, \sigma} (x ) 
       \end{array} \right] 
\;\;\;\; , 
\label{genk.4}
\eneq
\noindent
with $U$ defined so that
\be 
U
\vec t_R= t_R \left[ \begin{array}{c}1\\0  \\0\\ \vdots \\0
\end{array}\right] \:\: , \: \  
U\vec t_L=  t_L e^{i\delta}
\left[\begin{array}{c}\cos \theta \\ \sin \theta  \\0\\ \vdots \\0
\end{array}\right] \label{Utrans}
\:\:\:\: . 
\ee
\noindent
The phases, $\delta$ and $\theta$ are determined by the scalar product of the two vectors:

\be 
\vec t_R^*\cdot \vec t_L=t_L  t_R e^{i\delta}\cos \theta 
\label{thea}
\:\:\:\: .\ee
\noindent
The first phase $\delta$ can be adsorbed into $\chi$, the phase difference of the order parameters in the two leads.  
Thus, for a generic number of equivalent channels $K$, we can 
simply work with a $K=2$-model with real tunnelling matrix elements given by the 
right hand sides of Eq. (\ref{Utrans}), with $\delta =0$. Therefore, with 
no loss of generality, from now on we will assume $K=2$. For $K=2$,  
the general form of  $R_L ( E ; \chi)$ and  $R_R ( E ; \chi)$  
may be inferred by noting that the allowed 
physical processes at each interface are the ones corresponding to a 
particle (hole) incoming with spin $\sigma$ from channel 1 (2) and emerging  as a particle (hole) with spin $\sigma$ 
in channel 1 (2) after a normal reflection process, or as a hole (particle) with spin $- \sigma$ in channel 1 (2)
after an Andreev reflection process. 
In addition, there will be inter-channel reflection processes, in which a
particle (hole) incoming with spin $\sigma$ from channel 1 (2) can emerge as a particle (hole) with spin $\sigma$
in channel 2 (1) after a normal reflection process, or as a hole (particle) with spin $- \sigma$ in channel 2 (1)
after an Andreev reflection process. For notational simplicity, when dealing with the $K=2$-problem, in the remainder of this appendix and 
in next one, we will order the $ [ R_{L(R)} ( E ; \chi) ]_{(a, \rho) , (a' , \rho')}$-matrix elements in 
square matrices ${\cal R}_{L(R)} ( E ; \chi)$, so that,  denoting with $N^{p(h)}_{L(R) ,  ( \rho , \rho')} ( E ; \chi) $ 
and with $A^{p(h)}_{L(R) , (\rho , \rho')} ( E ; \chi) $
the single-particle(hole) normal and Andreev scattering amplitude at the left-(right-)hand interface 
from channel-$\rho'$ to  channel-$\rho$ respectively,  the matrices 
${\cal R}_{L(R)} ( E ; \chi) $  are  given by  (dropping for simplicity the arguments $E$ and 
$\chi$ from the matrix elements) 

\beq
{\cal R}_{L(R)} ( E ; \chi) = \left[ \begin{array}{cccc}
N^p_{L(R), (1,1)}  &   A^p_{L(R), (1,1)}  & N^p_{L(R), (1,2)}   &  A^p_{L(R), (1,2)}    \\
A^h_{L(R), (1,1)}   &   N^h_{L(R), (1,1)}   & A^h_{L(R), (1,2)}   &  N^h_{L(R), (1,2)}    \\
N^p_{L(R), (2,1)}   &   A^p_{L(R), (2,1)}   & N^p_{L(R), (2,2)}   &  A^p_{L(R), (2,2)}    \\
A^h_{L(R), (2,1)}   &   N^h_{L(R), (2,1)}   & A^h_{L(R), (2,2)}   &  N^h_{L(R), (2,2)}  
                        \end{array} \right]
\:\:\:\: . 
\label{ss.4}
\eneq
\noindent
At the Fermi level, Eq. (\ref{ss.4}) yields the matrices $\bar{{\cal R}}_{L(R)} ( \chi ) $, defined as 
 
\beq
\bar{{\cal R}}_{L(R)} ( \chi ) \equiv {\cal R}_{L(R)} ( E = 0  ; \chi ) =     \left[ \begin{array}{cccc}
\bar{N}^p_{L(R), (1,1)}    &   \bar{A}^p_{L(R), (1,1)}    & \bar{N}^p_{L(R), (1,2)}    
&  \bar{A}^p_{L(R), (1,2)}       \\
\bar{A}^h_{L(R), (1,1)}     &   \bar{N}^h_{L(R), (1,1)}    & \bar{A}^h_{L(R), (1,2)}     & 
\bar{N}^h_{L(R), (1,2)}      \\
\bar{N}^p_{L(R), (2,1)}    &   \bar{A}^p_{L(R), (2,1)}    & \bar{N}^p_{L(R), (2,2)}    
&  \bar{A}^p_{L(R), (2,2)}       \\
\bar{A}^h_{L(R), (2,1)}    &   \bar{N}^h_{L(R), (2,1)}    & \bar{A}^h_{L(R), (2,2)}     
&  \bar{N}^h_{L(R), (2,2)}    
                        \end{array} \right]
\:\:\:\: ,
\label{ss.5}
\eneq
\noindent
with the bar generically used to denote quantities evaluated at the Fermi level. By virtue of 
the charge-conjugation symmetry of the Bogoliubov - de Gennes equations, one 
finds that the following relations hold for the reflection amplitudes at 
the Fermi level:

\beq
\bar{N}^p_{L(R); ( \rho , \rho' )}   = [ \bar{N}^h_{L(R); ( \rho , \rho' )} ]^*    \;\;\; , \;\;
\bar{A}^p_{L(R); ( \rho , \rho' )}    =   [ \bar{A}^h_{L(R); ( \rho , \rho' )} ]^*  
\:\:\:\: . 
\label{ss.6}
\eneq
\noindent
As a result, after dropping the indices $^p$ and $^h$ and setting $\bar{N}_{L(R), (\rho,\rho')} 
\equiv  \bar{N}^p_{L(R), (\rho , \rho')}$ and  $\bar{A}_{L(R), (\rho,\rho')} 
\equiv  \bar{A}^p_{L(R), (\rho , \rho')}$, 
Eq. (\ref{ss.5}) can be rewritten as

\beq
\bar{{\cal R}}_{L(R)}  ( \chi ) =     \left[ \begin{array}{cccc}
\bar{N}_{L(R), (1,1)}   &   \bar{A}_{L(R), (1,1)}   & \bar{N}_{L(R), (1,2)}   &  \bar{A}_{L(R), (1,2)}    \\
\: [ \bar{A}_{L(R), (1,1)} ]^*   &   [ \bar{N}_{L(R), (1,1)} ]^*   &   [ \bar{A}_{L(R), (1,2)} ]^*   &  
[ \bar{N}_{L(R), (1,2)} ]^*  \\
\bar{N}_{L(R), (2,1)}   &   \bar{A}_{L(R), (2,1)}   & \bar{N}_{L(R), (2,2)}   &  \bar{A}_{L(R), (2,2)}    \\
\: [ \bar{A}_{L(R), (2,1)} ]^*   &   [ \bar{N}_{L(R), (2,1)} ]^*  &  [ \bar{A}_{L(R), (2,2)}]^*   &  
[\bar{N}_{L(R), (2,2)} ]^*  
                        \end{array} \right]
\:\:\:\: .
\label{ss.7}
\eneq
\noindent
Let us, now, compute ${\cal P} ( u ; \chi )$. Consistently with 
Eq. (\ref{thea}), we assume  

\beq
[ t_L ]_1 = t_L \cos ( \theta ) \; , \;
[ t_L ]_2 = t_L \sin ( \theta ) \;\;\; , \;\;
[ t_R ]_1 = t_R   \; , \;
[ t_R ]_2 =0
\:\:\:\: . 
\label{sll.2}
\eneq
\noindent
For the sake of computing ${\cal P} ( u ; \chi )$, it is useful to use the equivalence between
the electronic channels within C to rotate $\Psi_{C,1, \sigma} ( x )  , \Psi_{C , 2 , \sigma} ( x ) $ to 
$\bar{\Psi}_{C , 1, \sigma} ( x )  , \bar{\Psi}_{C, 2 , \sigma} ( x ) $, 
defined as

\beq
\left[ \begin{array}{c}
\bar{\Psi}_{C,1, \sigma} ( x )  \\ \bar{\Psi}_{C, 2 , \sigma} ( x )          
       \end{array} \right] = \left[ \begin{array}{cc}
\cos ( \theta  ) & \sin ( \theta  ) \\ - \sin ( \theta  ) & \cos ( \theta  )                                      
                                    \end{array} \right]
 \left[ \begin{array}{c}
\Psi_{C , 1 , \sigma} ( x )  \\ \Psi_{C , 2 , \sigma} ( x )          
       \end{array} \right]                                    
\:\:\:\: . 
\label{sll.3}
\eneq
\noindent
 Clearly, at the left(right)-hand SN interface, $\bar{\Psi}_{C,2,\sigma} ( x ) $ ($\tilde{\Psi}_{C,2,\sigma} ( x ) $) is
fully decoupled from the superconducting lead and can only exhibit normal reflection
at the Fermi level. As a result, in the basis of the operators $\bar{\Psi}_{C,1,\sigma} ( x )  , 
\bar{\Psi}_{C,2,\sigma} ( x ) $, 
one finds 

\beq
\bar{{\cal R}}_L = \left[ \begin{array}{cc} \bar{R}_L^{(1)} & {\bf 0 } \\ {\bf 0 } & 
                    - {\bf I}
                   \end{array} \right]
\:\:\:\: , 
\label{sll.5}
\eneq
\noindent
with $\bar{R}_L^{(1)}$ being the ($2 \times 2$) backscattering matrix for channel 1
at the left-hand interface, 
evaluated at the Fermi level.
Similarly,  in the basis of the operators $\Psi_{C,1,\sigma} ( x )  , 
\Psi_{C,2,\sigma} ( x )  $, 
one finds 

\beq
\bar{{\cal R}}_R = \left[ \begin{array}{cc} \bar{R}_R^{(1)} & {\bf 0 } \\ {\bf 0 } & 
                    - {\bf I}
                   \end{array} \right]
\:\:\:\: , 
\label{sll.7}
\eneq
\noindent
with $\bar{R}_R^{(1)}$ being the ($2 \times 2$) backscattering matrix for channel 1 at the 
right-hand interface, evaluated at the Fermi level.
 Taking into account
the need for rotating back and forth from the original basis $ ( \Psi_{C,1,\sigma} ( x )  , 
\Psi_{C,2,\sigma} ( x )  )$ 
to the   basis $ ( \bar{\Psi}_{C,1,\sigma} ( x )  , \bar{\Psi}_{C,2,\sigma} ( x )  )$,
in which the matrices $\bar{{\cal R}}_R$ and  $\bar{{\cal R}}_L $ are 
respectively block-diagonal, one finds that Eq. (\ref{sol.9}) yields 
 
\begin{eqnarray}
&& {\cal P} ( u ; \chi ) = {\rm det} \Biggl\{ {\bf I}_4 u- \nonumber \\
&& \left[ \begin{array}{cc}
\bar{R}_L^{(1)} & {\bf 0} \\ {\bf 0} & - {\bf I}                                                         
                                                       \end{array} \right]
\cdot \left[ \begin{array}{cc} e^{ i \sigma^z \alpha_F \ell}  & {\bf 0 } \\ {\bf 0} & e^{ i \sigma^z \alpha_F \ell} \end{array} \right] 
\cdot \left[ \begin{array}{cc}
\cos ( \theta ) {\bf I} & - \sin ( \theta )     {\bf I} \\   \sin ( \theta )     {\bf I}
& \cos ( \theta ) {\bf I}
             \end{array} \right] \cdot  \left[ \begin{array}{cc}
\bar{R}_R^{(1)} & {\bf 0} \\ {\bf 0} & -  {\bf I}                                                         
                                                       \end{array} \right]
 \cdot \left[ \begin{array}{cc} e^{ i \sigma^z \alpha_F \ell} & {\bf 0 } \\ {\bf 0} & e^{ i \sigma^z \alpha_F \ell} \end{array} \right] 
 \cdot \left[ \begin{array}{cc}
\cos ( \theta ) {\bf I} &  \sin ( \theta )     {\bf I} \\  -   \sin ( \theta )     {\bf I}
& \cos ( \theta ) {\bf I}
             \end{array} \right]
 \Biggr\}
 \nonumber \\
 &&=  {\rm det} \Biggl\{ {\bf I}_4 u- \nonumber \\
&& \left[ \begin{array}{cc} \cos^2 ( \theta ) 
\bar{R}_L^{(1)} \cdot e^{ i \sigma^z \alpha_F \ell} \cdot \bar{R}_R^{(1)} \cdot e^{ i \sigma^z \alpha_F \ell} - 
\sin^2 ( \theta ) \bar{R}_L^{(1)} \cdot  e^{ 2 i \sigma^z \alpha_F \ell}  & \cos ( \theta ) \sin ( \theta ) 
[ \bar{R}_L^{(1)} \cdot e^{ i \sigma^z \alpha_F \ell}
 \cdot \bar{R}_R^{(1)} \cdot e^{ i \sigma^z \alpha_F \ell} + \bar{R}_L^{(1)} \cdot   e^{2 i \sigma^z \alpha_F \ell}   ] \\
 \cos ( \theta ) \sin ( \theta ) 
[ -  e^{ i \sigma^z \alpha_F \ell} \cdot \bar{R}_R^{(1)} \cdot e^{ i \sigma^z \alpha_F \ell}-   e^{ 2 i \sigma^z \alpha_F \ell}  ]  & 
\cos^2 ( \theta )   e^{ 2 i \sigma^z \alpha_F \ell}  - \sin^2 ( \theta ) e^{ i \sigma^z \alpha_F \ell} \cdot \bar{R}_R^{(1)} \cdot 
e^{ i \sigma^z \alpha_F \ell}
\end{array} \right] \Biggr\} \nonumber \\ &&
\label{sll.11}
\end{eqnarray}
\noindent
with the suffix $_\theta$ added to ${\cal P} ( u ; \chi )$ to explicitly evidence
its dependence on $\theta$. 
As a consistency check of Eq. (\ref{sll.11}), we notice that, as $\theta \to 0$, we obtain 

\beq
{\cal P}_{\theta = 0 } ( u ; \chi ) = ( u - e^{ 2 i \alpha_F \ell} ) ( u - e^{ - 2 i \alpha_F \ell} ) 
P_2 ( u ; \chi )
\;\;\;\; , 
\label{sll.12}
\eneq
\noindent
with

\beq
P_2 ( u ; \chi ) = 
u^2 - 2 u \hbox{ Re } \{ e^{ 2 i\alpha_F \ell} \bar{N}_{L , (1,1)}^p \bar{N}_{R , (1,1)}^p + 
\bar{A}_{L , (1,1)}^h \bar{A}_{R , (1,1)}^p \} + 1  
\:\:\:\: . 
\label{fin.2}
\eneq
\noindent
Clearly, the only roots of ${\cal P}_{\theta = 0} ( u  ; \chi  ) $ that depend on  $\chi$, $u_\pm ( \chi )$, 
are the solutions of $P_2 ( u ; \chi ) = 0$. Setting $u_\pm ( \chi ) = e^{ \pm i \vartheta ( \chi ) }$ and 
using Eqs.(\ref{sol.12}), one then finds the main result of the
derivation of [\onlinecite{giu_af}], that is

\beq
I [ \chi ]  = - \frac{  e v}{ \pi \ell} \: \partial_\chi \vartheta^2 ( \chi )
\;\;\,\; . 
\label{fin.3}
\eneq
\noindent
This is definitely consistent with  Eq. (\ref{sll.12}) being the extension of 
the result of Eqs.(3,4) of Ref. [\onlinecite{giu_af}] to the case of a generic angle
$\theta$ between the couplings at the two SN interfaces. In the case we discuss at
the end of Section \ref{long_junction}, that is,  
two interfaces exhibiting perfect Andreev reflection, but with non-symmetric
couplings between the interfaces, that is, with $\theta \neq 0$,  one gets

\beq
\bar{R}_L^{(1)} = \left[ \begin{array}{cc} 0 & e^{ - \frac{i}{2} \chi} \\
        e^{   \frac{i}{2} \chi} & 0 
       \end{array} \right] \;\;\; , \;\;
\bar{R}_R^{(1)} = \left[ \begin{array}{cc} 0 &   e^{   \frac{i}{2} \chi} \\
          e^{ -   \frac{i}{2} \chi} & 0 
       \end{array} \right]
\;\;\;\; . 
\label{dc.1.5}
\eneq
\noindent
Insering the matrices $\bar{R}_L^{(1)} , \bar{R}_R^{(1)}$ into Eq. (\ref{sll.11}),
one obtains the polynomial ${\cal P}_\theta ( u ; \chi )$ in Eq. (\ref{dc.1.6}) of the main text.

\section{Construction of the function $\Phi [ \omega ; \chi ]$}
\label{phi_omchi}

In this appendix we develop a tecnique to derive the function
$\Phi [ \omega ; \chi ]$ defined in Section \ref{long_junctioni}, similar to the one we
use in Appendix \ref{co_put} to construct the polynomial ${\cal P} ( u )$. 
Moreover, we show how the main formula of Ref. [\onlinecite{gala_zaikin}] for
the zero-temperature dc Josephson current across 
a SINIS junction with $K$ channels within C can be recovered as a particular
limit of our results. The starting point is Eq. (\ref{ssol.3}) of Subsection \ref{co_put}, 
which we now develop without eventually imposing the symmetry constraint. 
On expanding the momenta analytically continued to imaginary energies we have now
to take into account the explicit dependence of the Fermi velocities on $\rho$, which 
yields  

\begin{eqnarray}
 \alpha^1_\rho &\approx& \alpha_F - i \frac{\omega}{v^{(\rho)}} \nonumber \\
 \alpha^2_\rho &\approx&  - \alpha_F - i \frac{\omega}{v^{(\rho)}} 
\:\:\:\: , 
\label{novel.1}
\end{eqnarray}
 \noindent
with $v^{(\rho)}$ being the Fermi velocity in channel-$\rho$, as defined 
after Eq. (\ref{ft.4}). Therefore, in the large-$\ell$ limit, 
Eqs.(\ref{novel.1}) motivate substituting in Eq. (\ref{ssol.3})
$e^{ i \alpha^a_\rho \ell}$ with $[ e^{ i \sigma^z \alpha_F \ell} ]_{a , a } 
e^{ - \frac{ \omega \ell}{v^{(\rho)} } }$. Moreover, just as we have done in 
the derivation in the symmetric case outlined in Appendix \ref{co_put}, 
we  set  $E=0$ in
the  matrices  $R_L ( E ; \chi)$ and  $R_R ( E ; \chi)$ and in the 
quasiparticle velocities $v^C_{a,\rho}$, which implies 
$\sqrt{ \frac{v_{a , \rho'}^C}{v_{a' , \rho}^C} } 
\approx \sqrt{ \frac{v^{ (\rho')}}{v^{(\rho)}} }$. 
As a result, one sees that, in   the large-$\ell$  limit, Eq. (\ref{ssol.3})
can be approximated as

\beq
{\rm det} \parallel  \delta_{a , a'} \delta_{ \rho , \rho'} - \sqrt{ \frac{v^{ (\rho')}}{v^{(\rho)}} }
( e^{ i \sigma^z \alpha_{F , \rho} \ell} )_{a,a} e^{ - \frac{\omega \ell}{v^{(\rho)}}}
\sum_{ a'' = 1 ,2 }\sum_{ \rho'' = 1}^K
[ R_R ( 0 ; \chi ) ]_{ ( a , \rho ) , (a'' , \rho'' ) }
( e^{ i \sigma^z \alpha_{F , \rho''} \ell} )_{a'' , a''}  e^{ - \frac{\omega \ell}{v^{(\rho'')}}} 
[ R_L ( 0; \chi ) ]_{ ( a'' , \rho'' ) , (a' , \rho' ) } \parallel = 0 
\;\;\;\; , 
\label{ssol.a3}
\eneq
\noindent
From the definition of the function $\Phi [ \omega ; \chi ]$ we give
in Eq. (\ref{newy.1}), we see that, once regarded as an equation in $\omega$ at
fixed $\chi$, Eq. (\ref{ssol.a3}) must have the same solutions as the 
equation $\Phi [ \omega ; \chi ] = 0$. Therefore, apart from an over-all multiplicative
nonzero coefficient, we obtain that $\Phi [ \omega ; \chi ]$ must coincide with the 
left-hand side of Eq. (\ref{ssol.a3}). By direct investigation, one finds that 
the appropriate multiplicative factor is given by $\prod_{ \rho = 1}^K e^{ \frac{ \omega \ell}{v^{(\rho)}}  }$. Thus, 
one eventually obtains

\begin{eqnarray}
 \Phi [ \omega ; \chi ] &=& 
\prod_{ \rho = 1}^K \{
\sum_{ \{ a_\rho , b_\rho \} \in \{ - 1 , 0 , 1 \} } [ [   \delta_{a_\rho , 0 } \delta_{b_\rho , 0 } + 
 \delta_{ | a_\rho | , 1 } \delta_{| b_\rho | , 1 } ]  e^{ i ( a_\rho - b_\rho )  \alpha_F^{(\rho)}  \ell } 
e^{ - w_\rho  ( a_\rho + b_\rho ) \omega} ] 
\nonumber \\&\times& 
 \bar{G}_{\{ a_1 , b_1 , \ldots , a_K , b_K \} } (  \chi )  
\}
\nonumber \\
&=&    {\rm det} \parallel  \delta_{a , a'} \delta_{ \rho , \rho'}e^{   \frac{\omega \ell}{v^{(\rho)}}}  \nonumber \\
&&   -  \sqrt{ \frac{v^{ (\rho')}}{v^{(\rho)}} }( e^{ i \sigma^z \alpha_{F , \rho} \ell} )_{a,a} 
\sum_{ a'' = 1 ,2 }\sum_{ \rho'' = 1}^K
[ R_R ( 0 ; \chi ) ]_{ ( a , \rho ) , (a'' , \rho'' ) }
( e^{ i \sigma^z \alpha_{F , \rho''} \ell} )_{a'' , a''}  e^{ - \frac{\omega \ell}{v^{(\rho'')}}} 
[ R_L ( 0; \chi ) ]_{ ( a'' , \rho'' ) , (a' , \rho' ) } \parallel 
\:\:\:\: . \nonumber \\
\label{ft.8}
\end{eqnarray}
\noindent
As a simple model calculation, let us now compute $\Phi [ \omega ; \chi ]$
for   $N=1$ and $K = 2$ inequivalent channels within C. In particular, 
to simplify the derivation, we
choose $H_T$ as in Eq. (\ref{sss.b1}) with $K=2$, but setting $\theta_L = \theta_R = \varphi$ in 
Eq. (\ref{sll.2}).  As it happens in the example of 
Appendix \ref{co_put}, also here only a linear combination of the  operators for
the two channels within C couples to the leads. Let $\bar{R}^{(1)}_L , \bar{R}^{(1)}_R$ be 
the corresponding $2 \times 2$ reflection amplitude matrix at the left-hand side and at 
the right-hand side interface for the coupled channel, respectively. One then obtains (defining
the square matrices ${\cal R} (E ; \chi)$ just as we did in Appendix \ref{co_put})

\beq
{\cal R}_{ L (R)} ( 0 ; \chi )    =  \left[ \begin{array}{cc}
 \cos^2 ( \varphi ) \bar{R}_{ L (R)}^{(1)} + \sin^2 ( \varphi ) {\bf I}   &   \cos ( \varphi ) \sin ( \varphi ) 
( \bar{R}_{ L (R)}^{(1)} - {\bf I } ) w  \\   \cos ( \varphi ) \sin ( \varphi ) 
( \bar{R}_{ L (R)}^{(1)} - {\bf I } ) \frac{1}{w}   &   \sin^2 ( \varphi ) \bar{R}_{ L (R)}^{(1)} + \cos^2 ( \varphi ) {\bf I} 
                    \end{array} \right]              
\:\:\:\: ,
\label{fft.1}
\eneq
\noindent
with $w = \sqrt{\frac{v^{(2)}}{v^{(1)}}}$. It is, now, simple to check that, 
for $\varphi = 0$ or $\varphi = \frac{\pi}{2}$,
respectively setting $u = e^{ - \omega w}$ and $u = e^{ - \frac{\omega}{w}}$, 
Eq. (\ref{ft.8}) gives back the (second-order) polynomial ${\cal P} ( u ; \chi  ) $
for a single-channel, with Fermi velocity and Fermi momentum 
equal to $v^{(1 )} , \alpha_{F,1}$ and to 
$v^{(2 )} , \alpha_{F,2}$, respectively. The same result is clearly obtained for a generic 
value of $\varphi$, on setting 
$\alpha_{F , 1 } = \alpha_{F,2} \equiv \alpha_F$ and $v^{(1)} = v^{(2)} = v$, which implies 
$w = 1$. In general, once $\Phi [ \omega ; \chi ]$ computed with 
Eqs.(\ref{ft.8},\ref{fft.1}) is put into Eqs.(\ref{ft.4},\ref{ft.4bis}), 
one recovers  a simple and effective tool to 
compute $I  [ \chi ; T=0 ]$ and  $I [ \chi ; T ]$ for generic values of the parameters 
by means of pertinent numerical techniques, as we do at the end of Section 
\ref{long_junctioni} by assuming perfect Andreev reflection at both interfaces, 
that is, by assuming that $\bar{R}_L^{(1)}$  and $\bar{R}_R^{(1)}$ are 
the matrices given in Eq. (\ref{dc.1.5}).

As mentioned in the introduction, from Eqs.(\ref{ft.3},\ref{ft.4}) it is possible to 
recover the main result of Ref. [\onlinecite{gala_zaikin}] for the dc 
Josephson current in a multi-channel SINIS-junction. To do so, 
one has to assume that there are no scattering processes at the interfaces
between different channels within C. Formally, this means that both the 
$R_L (E ; \chi ) $ and the $R_R ( E ; \chi )$ matrices (and, consequently, 
the $\bar{R}_L, \bar{R}_R$ matrices) have to be diagonal in the channel
index $\rho$, that is

\beq
 [ R_{L (R)} ( E ; \chi  ) ]_{ (a ,\rho ) ,(a' ,  \rho') } = 
[ R_{L (R)}^\rho ]_{a,a' } ( E ; \chi  ) \: \delta_{ \rho , \rho' } 
  \:\:\:\: . 
  \label{gz.1}
\eneq
\noindent
Accordingly, Eq. (\ref{ft.8}) for $\Phi  [ \omega ; \chi ] $ simplifies to

\beq
\Phi [ \omega ; \chi ] = \prod_{ \rho = 1}^K \Phi_\rho [ \omega ; \chi ]
\;\;\;\; , 
\label{gz.2}
\eneq
\noindent
with 

\beq
\Phi_\rho [ \omega ; \chi ] = e^{ w_\rho \omega } \: {\rm det} \Biggl\{ 
{\bf I}_2 - R_L^\rho (0 ; \chi ) \cdot \left[ \begin{array}{cc}
e^{ i \alpha_F^{(\rho)} \ell} e^{ - w_\rho \omega }  & 0   \\ 0 &   e^{ - i \alpha_F^{(\rho)} \ell}
  e^{ -  w_\rho \omega }                           \end{array} \right]
  \cdot  R_R^\rho ( 0 ; \chi )  \cdot \left[ \begin{array}{cc}
e^{ i \alpha_F^{(\rho)} \ell} e^{ - w_\rho \omega }  & 0   \\ 0 &   e^{ - i \alpha_F^{(\rho)} \ell}
  e^{ -  w_\rho \omega }                           \end{array} \right] \Biggr\}
  \:\:\:\: . 
  \label{gz.3}
  \eneq
  \noindent
As a result, at finite $T$ $I [ \chi ; T ]$ can be
written as  

\beq
I [ \chi ; T ] =\sum_{ \rho = 1}^K \{ 
2 e T \sum_{ \nu = - \infty}^\infty \: \partial_\chi \Phi_\rho [ 
\omega_\nu ; \chi ] \}  \equiv \sum_{ \rho = 1}^K I_\rho [ \chi ; T ] 
\;\;\;\; . 
\label{gz.4}
\eneq
\noindent
Similarly, at $T=0$ one obtains 

\beq
I [ \chi ; T ] =\sum_{ \rho = 1}^K \left\{ \frac{2 e }{2 \pi} \frac{U}{\ell} \:
\int_{- \infty}^\infty \: d \omega \: \partial_\chi \Phi_\rho [ \omega ; \chi ] \right\}
\equiv \sum_{ \rho = 1}^K I_\rho [ \chi ; T = 0 ]
\:\:\:\: . 
\label{gz.5}
\eneq
\noindent
$I_\rho [ \chi ; T ]$ and $I_\rho [ \chi ; T = 0 ]$ are the current for
a single-channel SNS junction at finite $T$ and at $T=0$, respectively. 
They can be readily computed following the derivation of 
Ref. [\onlinecite{giu_af}]. To compare with the  result of 
Ref. [\onlinecite{gala_zaikin}], we then compute $I_\rho [ \chi ; T = 0 ]$,
which is given by \cite{giu_af}

\beq
I_\rho  [ \chi ; T=0 ] = - \frac{ e v^{(\rho)}}{\pi \ell} \partial_\chi \vartheta_\rho^2 ( \chi ) 
\:\:\:\: , 
\label{gz.6}
\eneq
\noindent
with  

\beq
\vartheta_\rho ( \chi ) = {\rm arccos} \{  \hbox{Re} [  \bar{N}_{R , \rho}^p 
\bar{N}_{L , \rho}^p 
e^{ 2 i \alpha_{F  }^{(\rho)}  \ell}  
+  \bar{A}_{R , \rho}^p \bar{A}_{L , \rho}^h   ] \}
\label{gz.7}
\:\:\:\: ,
 \eneq
\noindent
and $\bar{N}_{R/L, \rho}^{p/h} , \bar{A}_{R/L , \rho}^{p/h}$ respectively being the 
normal and the Andreev single-particle/hole reflection amplitudes 
within channel-$\rho$ at the right/left-hand S-N interface
 evaluated at the Fermi level only. It is now straightforward to check that 
Eqs.(\ref{gz.6},\ref{gz.7}) give back the result of Ref. [\onlinecite{gala_zaikin}]
for a $K$-channel SINIS junction provided that, for a generic channel $\rho$, one
first of all relates the reflection and the transmission coefficients at the left(right)-hand 
SIN-interface, respectively given by $B_{L , \rho} , D_{L , \rho}$ ($B_{R , \rho} , 
D_{R , \rho}$) to the modulus of the normal and Andreev reflection
coefficients, according to the equations

\begin{eqnarray}
 |\bar{A}_{R/L , \rho}^{p/h} | &=& \frac{D_{R/L , \rho}}{1 +B_{R/L , \rho} }
 \nonumber \\
  |\bar{N}_{R/L , \rho}^{p/h} | &=& \frac{2 \sqrt{ B_{R/L , \rho}}}{1 +B_{R/L , \rho} }
\:\:\:\: , 
\label{gz.8}
  \end{eqnarray}
\noindent
and indentifies the phase $\phi_\rho$ in Eq. (23) of Ref. [\onlinecite{gala_zaikin}]
with ${\rm arg} [ \bar{N}_{R , \rho}^p 
\bar{N}_{L , \rho}^p ]$.  It is therefore likely that, where the range of applicability of
our approach overlaps with the one of the approach based on Eilenberger equations, 
equivalent results are obtained. 
It would be interesting to check this point by repeating, for instance, 
the calculations of Refs.[\onlinecite{gala_zaikin_2},\onlinecite{gala_zaikin_3}] 
with our technique, but this goes beyond the scope of this work, which is 
mainly a presentation of our approach. It is important to recall that, as
already remarked before, our derivation is amenable for trading complicated
model Hamiltonians describing the whole SNS junctions for simple boundary models, which
is the key steps for treating 
Luttinger liquid interaction effects in the central region.

\end{document}